\documentclass[aps,twocolumn,showpacs,floatfix]{revtex4}

\usepackage{amsfonts}
\usepackage{amsmath}
\usepackage{amssymb}
\usepackage{graphicx}
\usepackage{color}
\usepackage{verbatim}
\usepackage{lipsum}
\usepackage{dsfont}
\usepackage[utf8]{inputenc}
\usepackage{physics}

\usepackage[export]{adjustbox}
\usepackage{dcolumn}
\usepackage{bm}
\usepackage[mathlines]{lineno}

\usepackage[T1]{fontenc}

\usepackage{hyperref}
\usepackage{makecell}
\usepackage{upgreek}

\begin{document}

\preprint{APS/123-QED}

\title{
NV microscopy of thermally controlled stresses caused by Cr$_2$O$_3$ thin films}


\author{A. Berzins$^{1,2}$}
\email{andris.berzins@lu.lv}
\author{J. Smits$^{1,2}$}
\author{A. Petruhins$^3$}
\author{R. Rimsa$^4$}
\author{G. Mozolevskis$^4$}
\author{M. Zubkins$^4$}
\author{I. Fescenko$^1$}

\affiliation{$^1$Laser Centre, University of Latvia, Latvia}
\affiliation{$^2$The University of New Mexico, Albuquerque, United States}
\affiliation{$^3$Materials Design Division, Department of Physics, Chemistry and Biology (IFM), Linkoping University, Sweden}
\affiliation{$^4$Institute of Solid State Physics, University of Latvia, Latvia}

\date{\today}

\begin{abstract}
Many modern applications, including quantum computing and quantum sensing, use substrate-film interfaces. Particularly, thin films of chromium or titanium and their oxides are commonly used to bind various structures, such as resonators, masks, or microwave antennas, to a diamond surface. Due to different thermal expansions of involved materials, such films and structures could produce significant stresses, which need to be measured or predicted. In this paper, we demonstrate imaging of stresses in the top layer of diamond with deposited structures of Cr$_2$O$_3$ at temperatures 19$^{\circ}$C and 37$^{\circ}$C by using stress-sensitive optically detected magnetic resonances (ODMR) in NV centers. We also calculated stresses in the diamond-film interface by using finite-element analysis and correlated them to measured ODMR frequency shifts. As predicted by the simulation, the measured high-contrast frequency-shift patterns are only due to thermal stresses, whose spin-stress coupling constant along the NV axis is 21$\pm$1 MHz/GPa, that is in agreement with constants previously obtained from single NV centers in diamond cantilever. We demonstrate that NV microscopy is a convenient platform for optically detecting and quantifying spatial distributions of stresses in diamond-based photonic devices with micrometer precision and propose thin films as a means for local application of temperature-controlled stresses. Our results also show that thin film structures produce significant stresses in diamond substrates, which should be accounted for in NV-based applications.

\end{abstract}

\maketitle


\section{Introduction}

The Nitrogen-Vacancy (NV) centers in diamonds are point defects consisting of a vacancy in the diamond lattice adjacent to a substitutional nitrogen atom~\cite{ashfold_nitrogen_2020}. Long coherence times of their electron and nuclear spins and the possibility of being initialized and read optically made them widely studied as potential qubits and quantum sensors. Intensive studies of NV centers in the last decade have led to a large variety of sensing applications~\cite{wu_diamond_2016,chipaux_nanodiamonds_2018,norman_novel_2020,fu_sensitive_2020,abe_tutorial_2018}, which benefit from nanometer resolution and room-temperature operation of the NV-based devices, as well as from low toxicity and mechanical or chemical durability of their diamond matrix. Mainly these applications exploit the high sensitivity of NV centers to magnetic fields via ground state Zeeman effect by using the Optically Detected Magnetic Resonance (ODMR) detection ~\cite{rondin_magnetometry_2014}, while an auxiliary sensitivity to electric fields caused by the temperature or strain in the diamond lattice is isolated by measurement designs or disregarded. 

Over the last decade, the research on NV sensors for detecting strains or associated stresses in diamonds has intensified, likely due to progress in nano and micro photonics~\cite{aharonovich_diamond_2011,schukraft_invited_2016,ruf_optically_2019}, surface acoustic wave usage for phonon coupling in NV-based micro and nano mechanical systems~\cite{golter_optomechanical_2016,labanowski_voltage-driven_2018}, as well as quantum computing~\cite{ashfold_nitrogen_2020}. The interaction of NV electronic spins with deformations of a diamond crystal is a principal part of many quantum information processing schemes. For example, strain gradients could encode the position information of each NV center in a shift of its optical transition frequency, making densely packed NV centers individually addressable~\cite{xu_quantum_2019,bersin_individual_2019}. It has also been suggested to use the interaction between the electronic configuration of the NV ground state and local strains to increase the applicability of electron spin qubits by allowing to drive spin transitions, both magnetically allowed and forbidden~\cite{udvarhelyi_spin-strain_2018}. 

In brief, strain is the deformation or displacement of material that results from applied stress, which is the force applied to a material divided by the material’s cross-sectional area. Both strains (dimensionless) and stresses of a crystal are tensors, related by Young’s modulus or elastic modulus, usually expressed in gigapascals (GPa). Quantitative quantum sensing of strains or stresses in the diamond relies on knowledge of spin-strain or spin-stress coupling constants, which relate them to the shift of NV spin energy sublevels expressed in frequencies of the ODMR. 

Strains appear inadvertently in various nanoscale devices, exacerbated by their structural or material diversity. For example, stains could be created by chromium films that are often used in nanofabrication due to their exceptional adhesion to the diamond surface and have a linear thermal expansion coefficient, which is many times greater than that of the diamond. The physical and chemical properties of the materials, such as electronic, optical, magnetic, phononic, and catalytic properties, can be dramatically alternated under the application of stresses~\cite{li_elastic_2014}, especially at the nanoscale~\cite{smith_tuning_2009}. Fortunately for NV-based applications, the NV centers could be used for quantitative characterization of local stresses~\cite{barson_nanomechanical_2017, barfuss_spin-stress_2019}, providing intrinsic and accurate sensing, whatever we want to mitigate or exploit the strains. Ensembles of NV centers, in turn, can be used for mapping of stresses in a shallow NV layer~\cite{kehayias_imaging_2019,broadway_microscopic_2019,trusheim_wide-field_2016,ho__2021,ho_probing_2020,hsieh_imaging_2019}. The latest study demonstrates high sensitivity ac magnetometry for three-dimensional strain mapping with micron-scale spatial resolution with two order-of-magnitude improvement in volume-normalized sensitivity~\cite{marshall_directional_2021}. Besides the material quality control, NV sensing is recently proposed for strain-based detection of dark matter~\cite{marshall_directional_2021,rajendran_method_2017}.

Even though the coupling between the NV spin and the diamond deformation has been previously studied in combined quantum-mechanical systems, several inconsistencies exist in the available literature. Thus, this type of coupling constants has been investigated by use of a diamond anvil~\cite{barson_nanomechanical_2017}, and then consequently used in imaging of a partial stress tensor caused by intrinsic diamond defects~\cite{kehayias_imaging_2019} and imaging of full stress tensor caused by various mechanical impacts~\cite{broadway_microscopic_2019}. The last paper also reported surprisingly large stress contributions from functional electronic devices fabricated on the diamond. Following these studies, a complete formalism was proposed~\cite{barfuss_spin-stress_2019} to describe the coupling between the NV spins and crystal deformation, which in combination with an experimental method based on single NV centers in a deformed diamond cantilever~\cite{teissier_strain_2014} leads to a new set of four spin-stress constants that differs by a factor of $\approx2$--3 from reported in Ref.~\cite{barson_nanomechanical_2017}. 

In this work we demonstrate a simple but accurate method of applying mechanical strains to a diamond crystal via Cr$_2$O$_3$ thin film microstructures by controlling the temperature of the film--diamond interface. The method benefits from different thermal expansion coefficients of the materials commonly used in NV sensing, producing stresses localized in a shallow layer close to the diamond surface that are homogeneous in planes parallel to the surface over relatively large regions under the film patches. On the one hand, we used widefield diamond microscopy to map the stresses as the strain-sensitive frequency shift of ODMR signals in a shallow NV layer. On the other hand, we calculated the stresses by finite-element methods based on our knowledge of the system geometry and involved material mechanical properties. The correlation of both methods allowed us to quantify spin-stress coupling in the NV coordinate basis and compare it to the recently reported range of values~\cite{barson_nanomechanical_2017, barfuss_spin-stress_2019}.

\section{Experimental methods}
\subsection{Detection principle}
\begin{figure*}[ht]
  \begin{center}
    \includegraphics[width=0.95\textwidth]{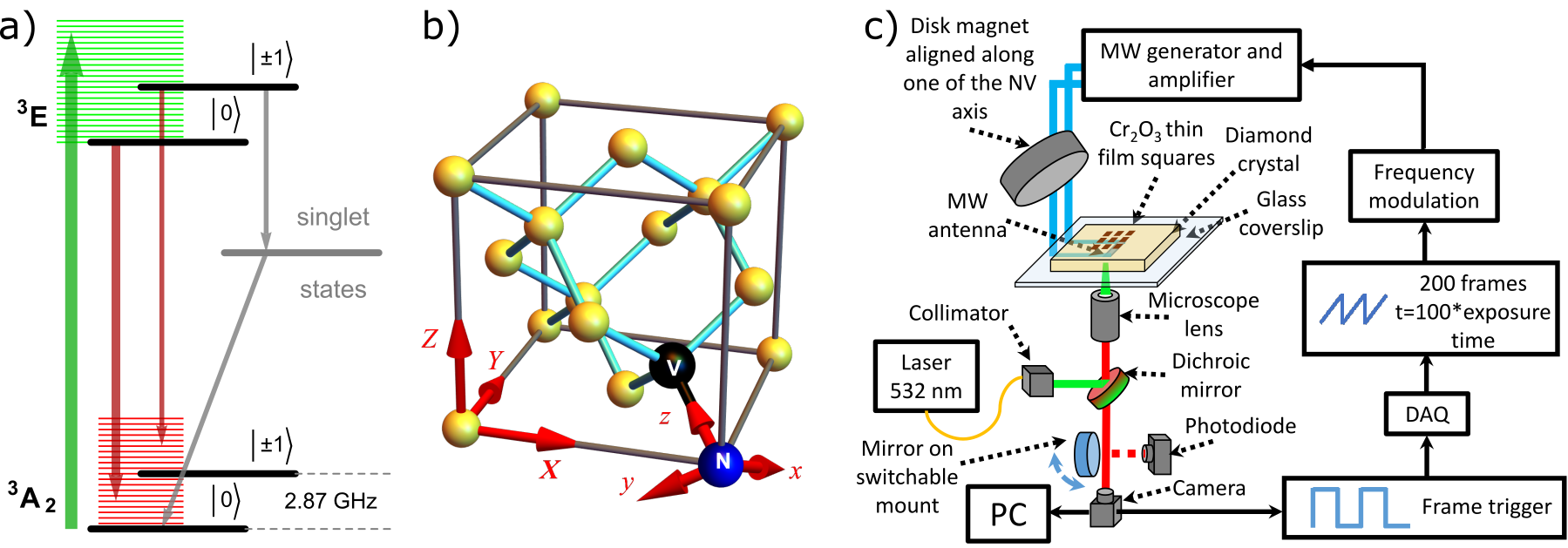}
  \end{center}
  \caption{\textbf{a)} NV energy level diagram depicting magnetic sublevels ($|0\rangle$, $|\pm1\rangle$), their phonon bands (green and red horizontal lines), optical excitation (green arrow), fluorescence (red arrows), and nonradiative (gray arrows) pathways. Gray arrows show spin-selective intersystem crossing leading to polarization into the \textbar0\big \rangle~ground-state sublevel. \textbf{b)} Graphical representation of the NV center in the diamond unit cell denoting the coordinate systems of the NV center ($xyz$) and the unit cell ($XYZ$) employed in this work. Yellow spheres depict carbon atoms, a blue sphere -- substitutional nitrogen, and a black sphere -- a lattice vacancy. \textbf{c)} Schematic of the experimental apparatus for widefield diamond microscopy. DAQ: data acquisition card; PC: the personal computer.}
  \label{1_2_3}
\end{figure*}

Figure \ref{1_2_3} a) depicts the NV energy levels with optical excitation and emission pathways. 
When the green laser light continuously excites NV centers, the ground-state population is optically polarized into the $|0\rangle$ state by a non-radiative, spin-selective decay via intermediate singlet states.
Because of the same spin-selective decay mechanism, NV centers excited from the $|0\rangle$ state emit fluorescence in the red region of the spectrum at a higher rate than those originating from $|\pm1\rangle$ states. Application of a transverse microwave (MW) magnetic field mixes the spin populations, resulting in a decrease in fluorescence (a resonance) when the MW frequency matches the spin transition frequencies, $f_{\pm}$. Evaluation of these frequencies by precisely measuring NV fluorescence as the MW frequency is swept across the resonances is the basic principle of ODMR techniques~\cite{rondin_magnetometry_2014}.

When an external magnetic field $B_{\parallel}>1$~mT is applied along one of the four possible N-V symmetry axis $z$, the $|0\rangle\leftrightarrow|\pm1\rangle$ spin transition frequencies are~\cite{kehayias_imaging_2019} 
\begin{equation}
\label{eqn:Freqs}
f_{\pm}=D(T)+M_z\pm\gamma_{NV}B_{\parallel},
\end{equation}
where $D(T)\approx2.87~{\rm GHz}$ is the temperature-dependent~\cite{acosta_temperature_2010} zero-field splitting at room temperature, $M_z$ is the effective electric field component associated with mechanical stress, and $\gamma_{NV}=28~{\rm GHz/T}$ is the NV gyromagnetic ratio. In general, there are four other components $M_x, M_e, N_x, N_y$ associated with mechanical stresses, which are significant only at $B_{\parallel}\approx0$~\cite{barfuss_spin-stress_2019,doherty_theory_2012,knauer_-situ_2020} or in the vicinities (51.2 and 102.4 mT) of levels anticrossing~\cite{wood_wide-band_2016,lazda_cross-relaxation_2021,busaite_dynamic_2020}.


In the simplest but full description, the coupling constant $M_z$ is expressed in terms of stress tensor components $\sigma_{ii}$ and spin-stress coupling constants (stress susceptibility parameters) $a_1$ and $a_2$~\cite{barfuss_spin-stress_2019} as
\begin{equation}
\label{eqn:Mz}
\begin{split}
    M_z &= a_1 (\sigma_{XX} + \sigma_{YY} + \sigma_{ZZ}) \\
    & + 2a_2 (\sigma_{YZ} + \sigma_{XZ} + \sigma_{XY}),
\end{split}
\end{equation}
where $XYZ$ are diamond unit cell reference coordinates and $xyz$ are NV reference coordinates (see Fig.~\ref{1_2_3} b)). One can see that the measurement described by Eqs.~\ref{eqn:Freqs} and~\ref{eqn:Mz} is only sensitive to the totals of axial and shear components of the stress tensor. A more sophisticated measurements are needed to obtain each individual $\sigma_{ii}$ contribution~\cite{kehayias_imaging_2019}. 
We empirically found in this study that stresses caused in a diamond by a film could be calculated by a truncated expression of Eq.~\ref{eqn:Mz} in the NV reference frame as
\begin{equation}
\label{eqn:aNV}
M_z \approx a_{NV} (\sigma_{xx} + \sigma_{yy} + \sigma_{zz}), 
\end{equation}
using a single spin-stress coupling constant $a_{NV}$ and axial components of the stress tensor. 

Typically, the detected frequency shift (see Eq.~\ref{eqn:Freqs}) is caused by simultaneous effects of the stresses in a hosting NV centers diamond crystals, the bias magnetic field, stray magnetic fields, and temperature shift. The magnetic field effects could be separated by measuring both frequencies $f_+$ and $f_-$ which are shifted by approximately equal amounts in opposite directions~\cite{fescenko_diamond_2019,fescenko_diamond_2020}. The shift of $f_+$ and $f_-$ by equal amounts in the same direction (common shift) is then caused by changes in local temperature and stresses. 
The NV sensing by widefield microscopy allows an alternative method for the separation of these effects based on their spatial distributions. The bias magnetic field and temperature (due to the exceptional thermal conductivity of diamond) produce homogeneous shift background, against which local stray fields~\cite{fescenko_diamond_2019} or longitudinal stresses could be imaged. When measuring at low bias magnetic fields $B_{\parallel}\sim10$mT, the resonance shifts due to paramagnetism/antiferromagnetism of the thin films are an order of magnitude smaller (see Fig.~\ref{cuts}) than those produced by the strain due to thermal expansion coefficient mismatch and thus can be neglected even when measuring only a single NV transition. 

\subsection{Apparatus}

The experimental setup for widefield imaging is depicted in Fig.~\ref{1_2_3} c). The diamond we use for the sensing is an electronics grade type IIa diamond (Element 6) with a (100) surface polish and with 3 mm $\times$ 3 mm $\times$ 0.1 mm dimensions. We performed \textbf{S}topping \textbf{R}ange of \textbf{I}ons in \textbf{M}atter or SRIM simulations~\cite{ziegler_srim_2010} to determine implantation parameters required for fabrication of 100-nm-thick NV layer close to the diamond surface. The SRIM simulation data for the diamond used can be seen in Ref.~\cite{berzins_characterization_2021}.
The crystal was irradiated with $^{14}$N ions at three different energies (10 keV, 35 keV, and 60 keV) with a cumulative dose of $5.5\times10^{12}~{\rm ions/cm}^2$ by CuttingEdge Ions, LLC and then annealed at 800$^{\circ}$C under high vacuum for four hours to promote migration of vacancies to substitutional nitrogen defects. 
 

The diamond sensor is placed on a coverslip in an epifluorescent microscope. The NV excitation and fluorescence detection are performed through the same infinity-corrected 40$\times$ microscope objective with a numerical aperture of 0.65 (Olympus Plan N). 
The NV centers are exposed to 100~mW radiation guided by a single mode optical fiber and lens system from a Coherent Verdi V-18 laser. A dynamic transmissive speckle reducer (Optotune LSR-3005) is inserted into the laser beam path to suppress interference artifacts originating from the thin air gap between the diamond and the coverslip, as well as within the diamond plate. The NV fluorescence ($650\mbox{--}800$ nm) is separated by a dichroic mirror (Thorlabs DMLP567R) and imaged through a long-pass filter (Thorlabs FEL0600) on an sCMOS sensor of Andor Neo 5.5 camera, or the photodiode (Thorlabs PDA36A-EC) for adjustments of the optical system. The field of view detected by the camera is a $110\times110~\upmu {\rm m}^2$. The air gap between the coverslip and the diamond reduces the diffraction-limited spatial resolution to $\sim 1.3~\upmu$m (NA of the microscope objective sets the fundamental resolution limit to $\sim 700$~nm).
 
The bias magnetic field $B_0\approx12.3$~mT is produced by a neodymium permanent disk magnet and aligned along one of the NV axes (surface polished along the (100) direction) at 35.5$^{\circ}$ to the diamond surface. 
The MW field is produced by an SRS SG384 generator, whose frequency is slowly swept by an analog voltage of a sawtooth waveform. The waveform is delivered from a data acquisition card (NI USB-6001). The amplified MW field (Mini Circuits ZHL-16W-43-S+ amplifier) is delivered by a copper antenna, which was e-beam deposited on the coverslip~\cite{fescenko_diamond_2019} in the form of a straight flat wire with $100\times1.5~\upmu \rm{m}$ cross-section. 

The MW frequency is swept across the central frequency of an ODMR ($|0\rangle\leftrightarrow|-1\rangle$) profile with a maximum deviation of $\pm$15 MHz. The MW sweep is triggered by a pulse from the sCMOS camera, followed by a series of 40 frames per sweep. One frame ($512\times512$ pixel) acquisition time is $90~\upmu$s. Five series are averaged in the camera memory before reading out by a LabVIEW interface. As a result, we obtain a series of NV fluorescence images revealing an ODMR shape for each pixel in a region of interest. In post-processing, we fit a series pixel-wise with a Lorentzian function to obtain a two-dimensional map of ODMR central frequencies. 

We present two series of ODMR images. One is measured at ambient temperature (19$^{\circ}$C) of the laboratory, and another is measured during heating the diamond at 37$^{\circ}$C.
When heating, Minco HK5207 thermofoil fed by Thorlabs TED200C temperature controller covered roughly half of the diamond plate, which had a thermocouple attached for temperature feedback. The low power of the laser and the MW fields do not contribute by a detectable amount to the heating of the diamond.

\subsection{Cr$_2$O$_3$ thin films}
\begin{figure*}
      \begin{center}
      \includegraphics[width=0.9\textwidth]{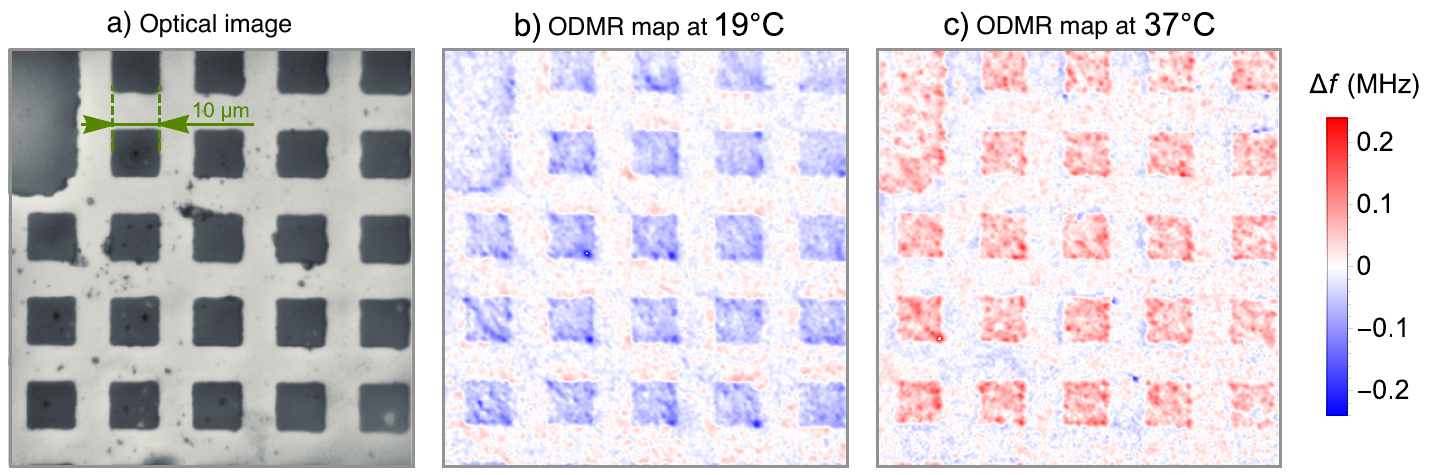}
    \end{center}
    \caption{\textbf{a)} The reflected light image of a grid of the 256~nm thick Cr$_2$O$_3$ squares is deposited on a diamond substrate. A structure in the left top corner is a defect of the grid, used as a spatial reference. \textbf{b)} and \textbf{c)} Maps of measured deviations $\Delta f$ from an averaged over uncoated areas value of ODMR $|0\rangle\leftrightarrow|-1\rangle$ frequency for 19~$^{\circ}$C and 37~$^{\circ}$C, respectively. The interrogated NV axes in the [100]-cut diamond substrate are aligned along a diagonal of the square and tilted 35.5$^{\circ}$ to the surface.}
  \label{maps}
\end{figure*}

Cr$_2$O$_3$ thin film is deposited by reactive pulsed-DC magnetron sputtering at 26~$^{\circ}$C (average temperature). The Cr target (99.95 \% purity) of 2" in diameter and 3 mm thick is kept at a 13-cm distance from the diamond substrate under the flow of Ar (20 sccm) and O$_2$ (3 sccm). The deposition is performed with an average power of 150 W at a chamber pressure of 1 Pa for 13 min. Following the film deposition, AZ 1518 photoresist is spin-coated for 30 s at 4000 RPM (1~s acceleration) and subsequently creates an etch mask by exposure via laser writer Heidelberg µPG101. Samples were immersed in a Chrome etchant (Sigma Aldrich 651826) for 35~s to remove Cr$_2$O$_3$ from exposed areas, followed by a rinse in deionized water. A grid of $10~\upmu$m~$\times~10~\upmu$m squares with a spacing of $10~\upmu$m between them is fabricated on the top of the NV sensor as a result (Fig.~\ref{maps}). The thickness of the squares is $256\pm1$~nm as is measured by the 3D optical profiler Zygo „NewView 7100”. Before the measurements, the diamond with the thin film structures was cleaned in Piranha solution (3:1 sulfuric acid and 30\% hydrogen peroxide solution) to remove any organic residue. 

Initially, the Cr$_2$O$_3$ structures were fabricated to study transitions from antiferromagnetic to paramagnetic properties at Néel temperature of 34~$^{\circ}$C. X-ray photoelectron spectroscopy (XPS) found that the thin films' chemical composition is only 67\% Cr$_2$O$_3$ and the remaining part is pure chromium -- a material with a seven-fold lower magnetic susceptibility. Considering the purpose of this study, that composition is not a disadvantage because both materials have the same excellent adhesion to the diamond surface and similar thermal expansion coefficients. The mean of linear thermal expansion coefficients reported in literature~\cite{alberts_elastic_1976,kirchner_thermal_1964,kudielka_thermische_1972} for Cr$_2$O$_3$ is $6.7\pm0.5\times10^{-6}$ K$^{-1}$ at room temperature. The corresponding coefficient of pure chromium is 6.6~$\times10^{-6}$ K$^{-1}$~\cite{hidnert_thermal_1941}. The linear thermal expansion coefficient of the diamond at room temperature is only 0.8~$\times10^{-6}$ K$^{-1}$~\cite{jacobson_thermal_2019}, which is eight times less than for Cr$_2$O$_3$.
This difference could lead to significant stresses in the diamond-chromium interface when the operating temperature is different than the film deposition temperature.

We calculate stresses in the diamond-chromium interface by finite element analysis using the solid mechanics package in the COMSOL Multiphysics platform. We perform a 3D simulation of nine chromium rectangular structures (each $10\times10\times0.25~\upmu$m$^3$), called here "squares" for simplicity, on the top of a diamond plate ($50\times50\times4~\upmu$m$^3$). The peripheral squares are necessary to correctly calculate stresses under the uncoated area around the central square. Coordinate axes of the geometry are directed along the diamond edges or such that one of the axes is along the NV axis, which is obtained by sequential rotation of 45$^{\circ}$ in the horizontal plane and 35.5$^{\circ}$ in the vertical plane relative to the diamond side.
Volume reference (stress-free) temperature is set to the deposition temperature (26~$^{\circ}$C).
Components of the stress tensors (Gauss point evaluation, spatial frame) are vertically averaged in a 200-nm-thick top layer of the diamond using the general projection function.
The same model is used with AC/DC package (mf physics) to calculate stray magnetic fields induced by the bias magnetic field. In this case we set magnetic susceptibility of the squares to $+1960~\times 10^{-6}$~$\text{cm}^3/\text{mol}$.

\section{Results and analysis}

\subsection{Frequency shift imaging}

Figure~\ref{maps} a) shows a reflected light image
of Cr$_2$O$_3$ thin films (dark) deposited on the diamond sensor’s surface (bright). 
This image is taken using the same optical system and camera, but the laser light is replaced with white light illumination. The small dark spots are likely the surface roughness of the diamond crystal. They do not appear in ODMR shift images shown in Fig.~\ref{maps} b) and c). However, the areas under the Cr$_2$O$_3$ structures are clearly frequency-shifted. Although the Cr$_2$O$_3$ films at 37~$^{\circ}$C are paramagnetic, we do not expect any magnetic signals to be detected at the applied bias magnetic field $B_0\approx12.3$~mT except the homogeneous shift $\pm\gamma_{NV}B_{\parallel}$. In a material with similar magnetic properties, such stray field structures are detected at hundreds of millitesla~\cite{fescenko_diamond_2019,berzins_surface_2021}.

Another parameter that causes a frequency shift is the temperature. The ODMR frequencies are homogeneously shifted by $D(T)$ with $74~{\rm kHz/K}$~\cite{acosta_temperature_2010}. The homogeneous background is subtracted by normalization to the mean frequency of uncoated areas of each image, revealing local deviations $\Delta f$ from the homogeneous shift. Patterns and signs of deviations $\Delta f$ are qualitatively consistent with expected changes in stress distributions due to cooling to 19~$^{\circ}$C and heating to 37~$^{\circ}$C of the film-diamond interface, which is stress-free at the deposition temperature of 26~$^{\circ}$C; because of the different thermal expansion coefficients, the cool films shorten NV axes in the upper layer of the diamond, but the warm films elongate them, resulting in corresponding changes of the effective electric field component $M_z$. 

\subsection{Stress calculation}

\begin{figure*}
\begin{center}
\includegraphics[width=0.9\textwidth,valign=t]{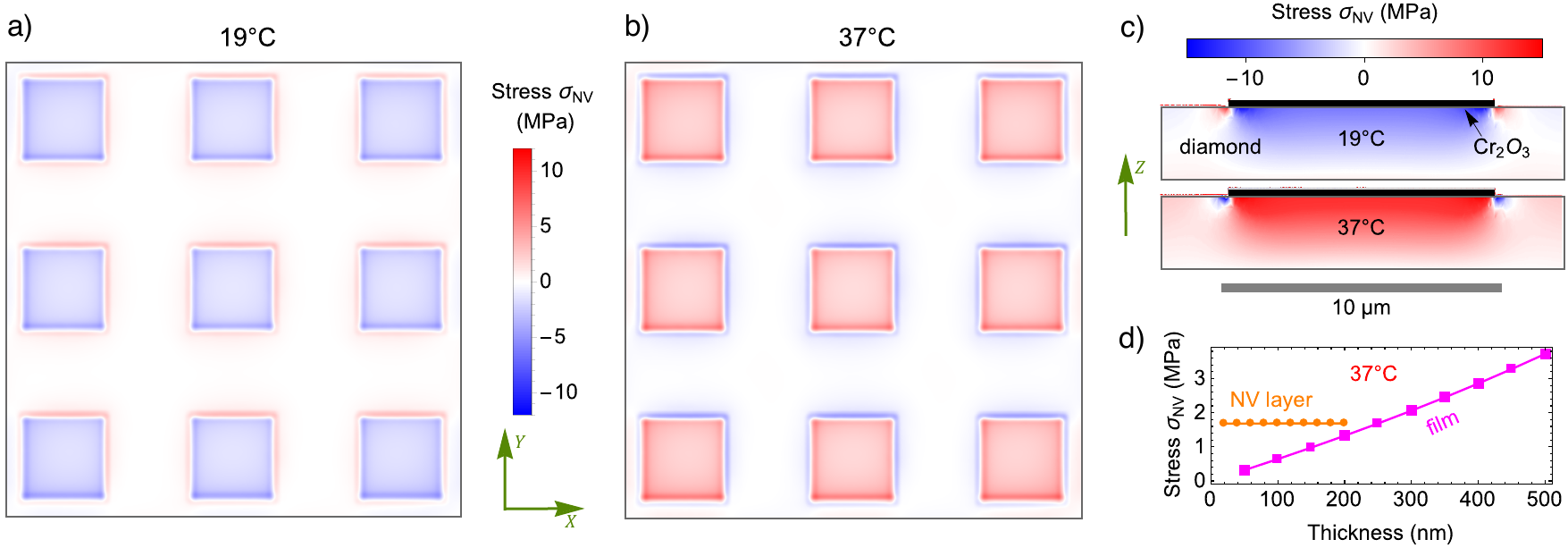}
\hspace{0.1cm}
\end{center}
\caption{Simulation of stresses $\sigma_{NV}=\sigma_{xx}+\sigma_{yy}+\sigma_{zz}$ averaged over a top 100-nm-thick layer of the diamond substrate at 19~$^{\circ}$C~\textbf{a)} and 37~$^{\circ}$C~\textbf{b)}. Stresses are caused by Cr$_2$O$_3$ thin structures on a diamond substrate at stress-free temperature of 26~$^{\circ}$C. Geometry and dimensions are the same as in Fig.~\ref{maps}. \textbf{c)} Simulated stresses in the XZ plane cut through the center of a Cr$_2$O$_3$ square.
The Cr$_2$O$_3$ films appear as "overexposed" (white or black colors) because they experience much larger stresses than the diamond under them. d) Stresses averaged over an NV layer under the center of a Cr$_2$O$_3$ square versus the thickness of the NV layer and the film. 
}
\label{Comsol}
\end{figure*}

The $M_z$ is usually expressed by all six unique components of the symmetric stress tensor in the crystal unit cell reference frame through two linear combination coefficients, see Eq.~\ref{eqn:Mz}. Therefore, a 2D map of any component of the sensor or totals of axial or shear components would be quantitatively different from a correlated experimental image~\cite{kehayias_imaging_2019}.
Here, we hypothesize that in the NV reference frame, the shear components are negligible, and the stresses could be conveniently expressed by a stress component $\sigma_{NV}=\sigma_{xx}+\sigma_{yy}+\sigma_{zz}$, which is related by a single conversion factor to the measured frequency shift $M_z$, see Eq.~\ref{eqn:aNV}. Figure~\ref{Comsol} shows the simulated patterns of stresses $\sigma_{NV}$ in the 100-nm thick top diamond layer, calculated with the procedure described in Sec. II. C. In the simulations, we account for the effect of optical diffraction by convolution with a two-dimensional Gaussian kernel (“blur”) with 700-nm FWHM. The simulations show that at 19~$^{\circ}$C Cr$_2$O$_3$ films squeeze atoms in the top layer of the diamond, and at 37~$^{\circ}$C, they stretch the atoms. These crystal lattice changes are almost homogeneous under the whole area of films except edges and qualitatively similar to the corresponding $\Delta f$ maps shown in Fig.~\ref{maps}. Near the edges of films, signs of stresses are flipped because the films push or pull the uncoated areas. For some reason, these edge effects are not prominent in the experimental maps. The difference between calculations and experiments might result from chemical etching-related artifacts that change the geometry of the thin film edge, diverging from a perpendicular side of the thin film square.

Figure~\ref{Comsol} c) shows the vertical distribution of stresses $\sigma_{NV}$ in the XZ plane cut through the center of a Cr$_2$O$_3$ square.
Figure~\ref{Comsol} d) shows vertically averaged $\sigma_{NV}$ under the center of Cr$_2$O$_3$ square versus the NV layer and the Cr$_2$O$_3$ film thickness. The NV layer thickness has a negligible impact on stresses $\sigma_{NV}$ under the film. Such a low impact is an advantage because the thickness of the NV layer is estimated indirectly from the simulation of ions implantations and diffusion of vacancies. In turn, the thickness of the film structures can be directly measured with nanometer precision, leading to a negligible contribution to stress uncertainty. The stress $\sigma_{NV}$ almost linearly depends on the film thickness providing an additional means to control stress in the NV layer. According to our simulation, there is an impact of the diamond plate dimensions on the stress magnitude. However, it becomes negligible when the thickness of the diamond substrate becomes larger than 20~$\upmu$m and its latitude larger than 50~$\upmu$m.

\begin{figure*}
      \begin{center}
   \includegraphics[width=0.9\textwidth,valign=t]{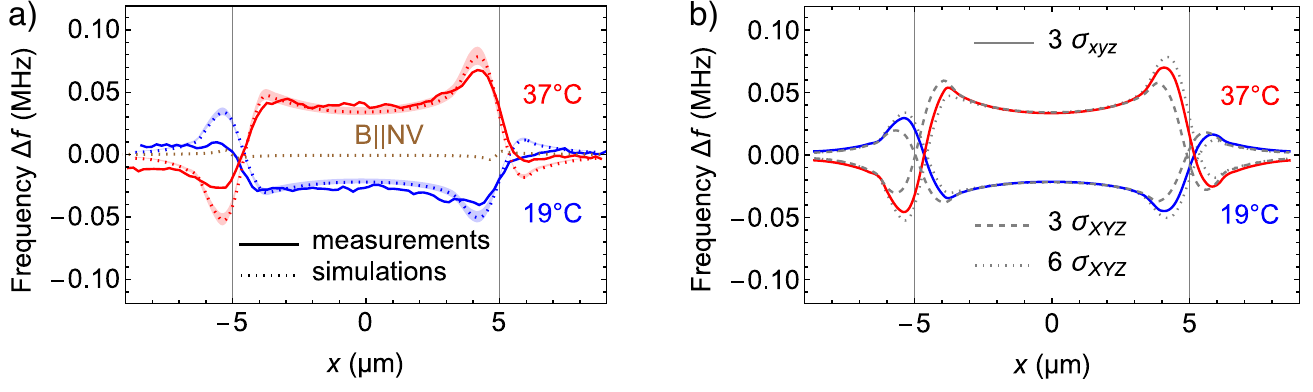}
          \end{center}
       \caption{\textbf{a)} Horizontal profiles of ODMR frequency maps obtained by averaging $\Delta f$ values over rows in an $\approx 4~\upmu$m wide band across a square formed by spatially overlaying and averaging 14 square features (see, Fig.~\ref{maps}). Solid lines show experimentally measured $\Delta f$, but dotted lines of the same color are $M_z$ calculated by finite-elements simulations with the help of Eq.~\ref{eqn:Mz} and constants reported in Ref.~\cite{barfuss_spin-stress_2019}. Confidence bands around the simulated curves are obtained by feeding into the model extreme estimations of mean thermal expansion coefficients for Cr$_2$O$_3$ crystals~\cite{alberts_elastic_1976,kirchner_thermal_1964,kudielka_thermische_1972}. Brown dotted lines represent the simulated paramagnetic signal created by the Cr$_2$O$_3$ film at 37~$^{\circ}$C. 
       \textbf{b)} Horizontal profiles of stress-induced $\Delta f$ maps calculated in the crystal coordinate system $XYZ$ with the use of the full stress tensor (six components of a symmetric tensor, dotted lines) and only diagonal stresses (three components, dashed lines) with spin-stress coupling constants $a_1$ and $a_2$ reported in~\cite{barfuss_spin-stress_2019}, and in $xyz$ coordinate system of the NV center (solid lines) with use of three diagonal components of the stress tensor multiplied by a factor of 18. 
       The range between the solid gray vertical lines marks the region where the Cr$_2$O$_3$ thin film is deposited on the diamond surface.}
  \label{cuts}
\end{figure*}

\begin{figure}[ht]
      \includegraphics[width=0.47\textwidth]{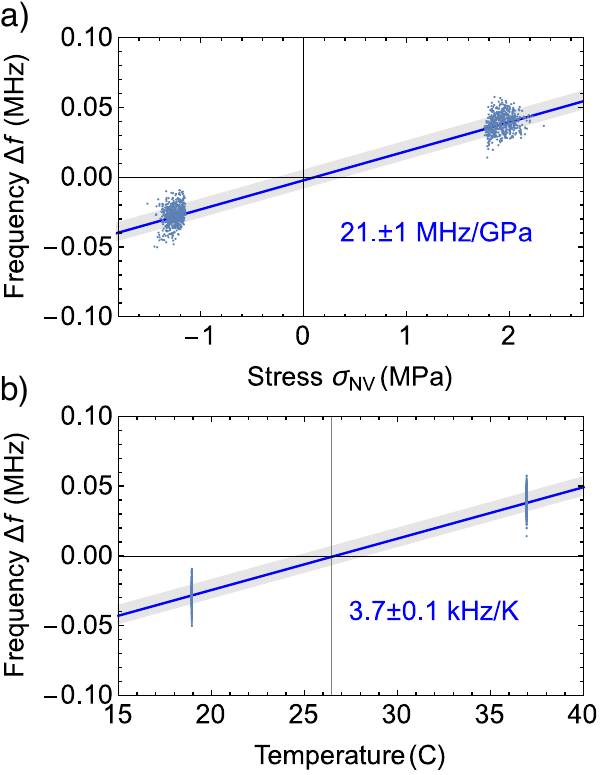}
    \caption{\textbf{a)} Measured frequencies $\Delta f$ (blue scattered dots) vs. simulated longitudinal stresses in the central region ($3\times3~\upmu {\rm m}^2$) of the Cr$_2$O$_3$ square. The longitudinal spin-stress coupling constant in the inset is derived from linear fit (solid blue line). The gray area depicts standard deviations of the fit residuals. \textbf{b)} The same measured frequencies $\Delta f$ as above but shown vs. temperature scale.}
  \label{sensitivity}
\end{figure}

\begin{figure}
             \includegraphics[width=0.46\textwidth,valign=t]{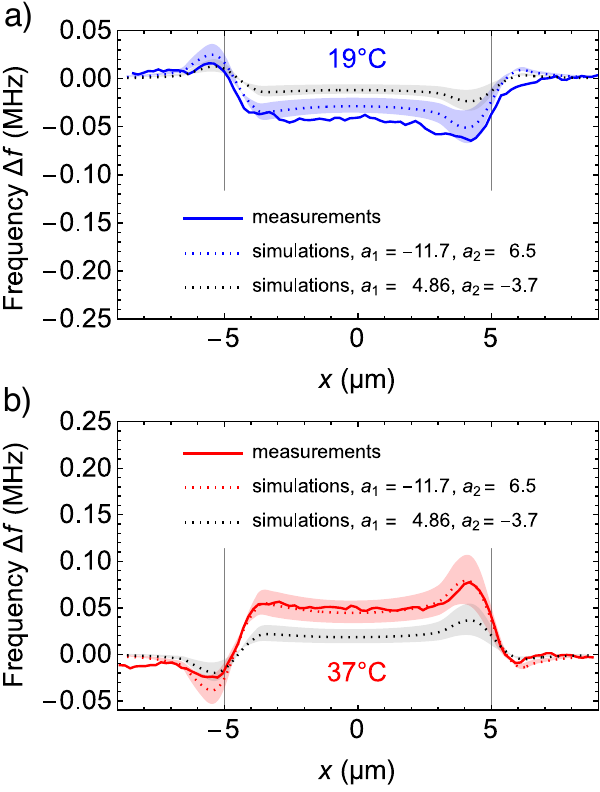}
             \caption{Horizontal profiles of measured maps of ODMR frequency $\Delta f$ for 19~$^{\circ}$C \textbf{a)} and 37~$^{\circ}$C \textbf{b)} compared to simulated NV frequencies by using spin-stress coupling constants reported in Ref.~\cite{barfuss_spin-stress_2019} (colored dotted lines) and in Ref.~\cite{barson_nanomechanical_2017} (gray dotted lines). Confidence bands are defined by uncertainties of the constants provided in the corresponding references. Note that compressive stress is defined to have negative amplitudes here, while Ref.~\cite{barson_nanomechanical_2017} defines positive amplitudes for it.
        }
  \label{BarBar}
\end{figure}

\subsection{Stress-spin coupling analysis}

To analyze the contribution of different effects and correlation of the measured and simulated data, we calculate profile plots from the 2D images. 
The measured maps are first prepared in the following way. To reduce the noise of $\Delta f$ images, we divide each image into segments with a thin film square in each and overlap them upon matching the squares. After taking a mean of the overlapped images, we obtain an averaged $\Delta f$ map with a single square of Cr$_2$O$_3$ film. Then the profiles are obtained by averaging $\Delta f$ values over rows ($x$ profile) in a $4~\upmu$m wide band across a square. The same profiles are calculated through the central square in calculated maps of stresses and maps of stray magnetic fields. In addition to the $\sigma_{NV}$ image, we calculate two maps of totals of axial and shear tensor components in the $XYZ$ reference coordinates and convert them into a map of $M_z$ by using Eq.~\ref{eqn:Mz} and coefficients from Ref.~\cite{barfuss_spin-stress_2019}.

Figure~\ref{cuts} a) shows the measured (solid lines) and calculated (dotted lines) profiles. The range between the solid gray vertical lines marks the diamond region coated with the Cr$_2$O$_3$ film.
The blue profiles are for measurements and simulations at the temperature of 19~$^{\circ}$C, but the red profiles are for the data at 37~$^{\circ}$C. Brown dotted lines depict the profiles of the calculated paramagnetic signals $B_\parallel$ from the Cr$_2$O$_3$ film at a temperature of 37~$^{\circ}$C. At the temperature of 19~$^{\circ}$C, the Cr$_2$O$_3$ film is antiferromagnetic, meaning that the stray magnetic field should not appear (thus omitted). Confidence bands around the simulated stress curves are obtained from extreme estimations of mean thermal expansion coefficients for Cr$_2$O$_3$ crystals~\cite{alberts_elastic_1976,kirchner_thermal_1964,kudielka_thermische_1972}. 

Both vertical and horizontal profiles show good qualitative and quantitative agreement between the experimental data and calculations with spin-stress coupling constants reported in Ref.~\cite{barfuss_spin-stress_2019}. The experimental data have less pronounced peaks near the edges of the films, likely because of the etching-related artifacts mentioned before and also due to imperfections in stacking images of 14 squares, which resulted in some blurring of sharp features. Nevertheless, the flips of the sign of stress are evident in the profiles. Moreover, experimental and simulated profiles have weak asymmetry due to the specific geometry where the NV axis is tilted relative to the surface. 
 
A question may arise if magnetic fields or temperature could also contribute to the shapes of the profiles. Regarding the first, we do not expect any magnetic features that could exceed the frequency noise at 37~$^{\circ}$C (see brown dotted lines in Fig.~\ref{cuts} a)) where the thin film is paramagnetic and even more so at 19~$^{\circ}$C where the thin film is antiferromagnetic. The temperature gradients are unlikely because of the excellent thermal conductivity of diamond. However, we could suppose that the laser radiation heats films on the diamond surface, leading to a slightly warmer NV layer volume under the Cr$_2$O$_3$ films. While we do not know an amplitude of such an alleged effect, our simulations with an arbitrary difference in diamond and film temperatures reveal that the temperature shift would be maximal under the center of a film structure, gently decreasing toward the edges of the film. The temperature-caused frequency shifts at both experimental temperatures (red and blue profiles) would be negative and would most of all affect the central parts of the profiles. As there is no evidence of such distortion in Fig.~\ref{cuts} a), we conclude that within the margin of error, only stress patterns contribute to the profile shapes and amplitudes.

\subsection{NV coordinate system}

In Fig.~\ref{cuts}~b) we compare profiles calculated in $XYZ$ reference frame with the help of precise Eq.~\ref{eqn:Mz} with profiles calculated in $xyz$ reference frame with the help of truncated Eq.~\ref{eqn:aNV}. The profiles are calculated using the same procedure as in Fig.~\ref{cuts}~a). One can see that both expressions lead to profiles, which are almost identical except for little differences in the vicinity of film edges. 
The $xyz$ reference frame is convenient for most NV-based measurements, and one has to know only one spin-stress coupling coefficient $a_{NV}$ to unambiguously link stresses experienced by NV axes with stress-induced frequency shifts. 
Using only axial components of Eq.~\ref{eqn:Mz} upon keeping the $XYZ$ reference frame could also predict stresses under the center of the film (dashed gray lines in Fig.~\ref{cuts}~b)), but become significantly wrong toward to edges of the film.

We estimate the coefficient $a_{NV}=21~\pm$~1~MHz/GPa from a linear fit of measured $\Delta f$ confined in the central region ($3\times3~\upmu {\rm m}^2$) of the Cr$_2$O$_3$ square versus corresponding totals ($\sigma_{NV}$) of axial components of the stress tensors in $xyz$ reference frame, as shown in Fig.~\ref{sensitivity} a). We obtain the temperature dependence of stress-induced frequency shift $d f/dT=3.7$~kHz/K, Fig.~\ref{sensitivity} b) from the same data set expressed in temperature units. This value is specific to the studied film-diamond interface and its magnitude is $\approx20\times$ less than the homogeneous temperature shift $d D/dT=74~{\rm kHz/K}$~\cite{acosta_temperature_2010}. However, a frequency shift of 3.7~kHz is comparable to an ODMR shift of 132~$\upmu$T in magnetic field measurements. Note that from Fig.~\ref{sensitivity} b) we can estimate the stress-free temperature, which is $\approx~26~^{\circ}$C -- in a reasonable agreement with the predefined temperature of the film deposition.

\subsection{Spin-stress coupling coefficients}

Two different sets of spin-stress coupling constants are recently reported~\cite{barfuss_spin-stress_2019,barson_nanomechanical_2017}. They differ by a factor $\sim2$--3 and in their signs. One set is obtained through applying uniaxial stress to a diamond cube in a diamond-anvil cell~\cite{barson_nanomechanical_2017}, and another set is obtained through applying stress to a diamond cantilever~\cite{barfuss_spin-stress_2019}. The weakness of both methods is the need to calculate local stresses using $macro$ dimensions of a whole diamond chunk. In this study, we calculated the local stresses based on well-known thermomechanical properties of the diamond and Cr$_2$O$_3$ thin film interface. Besides the mechanical properties, the local stresses in such a system depend only on the thickness of film structures, which is known with nanometer precision. On the other hand, the range of thermal linear expansion coefficients, which are reported in literature~\cite{alberts_elastic_1976,kirchner_thermal_1964,kudielka_thermische_1972}, leads to a narrow confidence band around calculated profiles of $\Delta f$ in Fig.~\ref{cuts}~a).

The relatively wider confidence bands in Fig.~\ref{BarBar} represent uncertainties of the spin-stress coupling coefficients reported in the literature~\cite{barfuss_spin-stress_2019,barson_nanomechanical_2017} for our calculated profiles of stress-induced $\Delta f$. These profiles are calculated using the same procedure as in Fig.~\ref{cuts}~a) and shown alongside the measured profiles. The constants $a_1$ and $a_2$ (Eq.~\ref{eqn:Mz}) used for gray profiles (dotted lines) are from Ref~\cite{barson_nanomechanical_2017}. This set of constants, which is often used in stress imaging~\cite{broadway_imaging_2020, kehayias_imaging_2019}, does not predict the strong frequency shift signals we observe. Profiles calculated with a set of constants $a_1$ and $a_2$ from Ref.~\cite{barfuss_spin-stress_2019},
are presented as blue and red dotted lines for 19~$^{\circ}$C and 37~$^{\circ}$C, respectively, and our measurements support spin-stress coupling coefficients from this set.

\section{Summary and outlook}

We have applied the widefield stress imaging technique previously demonstrated in Refs.~\cite{kehayias_imaging_2019, broadway_imaging_2020} to a diamond-film interface. In a proof-of-principle study, we demonstrated the ability of our method to estimate spin-stress coupling coefficients and confirmed coefficients previously found in Ref.~\cite{barfuss_spin-stress_2019}. Unlike other methods of spin-stress measurements, our method is not limited by the uncertainty of macroscopic dimensions of a diamond sample. The diamond-film interface is a versatile tool as it enables both: local stress sensing by ODMR technique and local stress-inducing utilizing ambient temperature.

There is a number of methods that can be used for measurements of stresses or associated strains in diamond: Rayleigh scattering~\cite{froggatt_high-spatial-resolution_1998}, electron spectroscopy ~\cite{cai_high-spatial-resolution_2005}, X-ray tomography and microscopy~\cite{marshall_scanning_2021,moore_imaging_2009,friel_control_2009}, Raman spectroscopy~\cite{crisci_residual_2011}, cathodoluminescence~\cite{p_l_hanley_i_kiflawi_and_andrew_richard_lang_topographically_1977}, and birefringence measurements~\cite{kehayias_imaging_2019,friel_control_2009,crisci_residual_2011}. However, these methods require the light or electrons to pass through the diamond, while the stress information is accumulated throughout the irradiated volume. The exception is atomic force microscopy and X-ray microscopy~\cite{kempf_high_2007}, which can only be used for surface measurements.
Unlike these methods, NV imaging can detect stresses in a hundred-nanometer thick layer fabricated next to the surface of the diamond. This technique has many degrees of freedom, which are still unexplored. There are four different directions of NV axes, and any of them can be interrogated separately. Besides, the diamond surface can be cut to create NV ensembles with axes parallel or normal to the surface. Finally, dispersed NV centers could be interrogated for stresses one by one, providing nanometer resolution in an imaging plane. Moreover, the optical detection of magnetic resonance allows image stresses through the diamond without accumulating the stress information outside the NV layer.

We used thin films of Cr$_2$O$_3$ to induce and alter stresses in the NV layer. Under the film at a distance of $\sim~2~\upmu$m from its edges, the magnitude of such stresses is homogeneous in a plane parallel to the diamond surface, and its uncertainty is limited by the thickness of the thin film, which is known with nanometer precision. The stress appears when the ambient temperature is different from the film deposition temperature due to the difference in thermal expansion coefficients. By setting the ambient temperature, one can apply desired stress to the NV centers, for example, to move it off and on magnetic resonance. 

Stress gradients created in normal to the diamond surface direction or near the edges of the film can encode the position information of each NV center in a shift of its optical transition frequency, making densely packed NV centers individually addressable~\cite{xu_quantum_2019,bersin_individual_2019}, thus specifically designed systems could surpass optical resolution without using super-resolution microscopy techniques. Additional studies are needed to estimate maximum stresses and their gradients achievable by the diamond-film interface. 

On the other hand, stresses could be mitigated by operating a device at a stress-free temperature. One example of such a device would be a microwave antenna that can be deposited on the surface of the diamond; this gives a well-controlled distance between the NV layer and the emitter~\cite{chipaux_magnetic_2015}. However, the antenna (typically 5 nm thick chromium adhesion layer coated with hundreds~nm thick gold or copper film) creates stress gradients near its edges that effectively change the orientation and length of NV axes as the strain bends the crystal. 
Moreover, when the microwave radiation flows through the copper wire, the electrical resistance could cause changing temperature, resulting in dynamically changing stress gradients.
This could lead to a degraded sensitivity of an NV ensemble, as the ODMR profiles will get wider, and as a consequence, NV centers near the antenna edges should not be used. 
Measuring stresses under the antenna and uncoated surface could reveal a stress-free temperature. 
Similar principles might apply to reflective coatings used for some applications and other cases of material deposition on the surface of the diamond~\cite{ukhina_influence_2020} as well as gluing diamond crystals to different surfaces~\cite{duan_efficient_2019,ho__2021}.

This study shows that a difference of just seven degrees from the stress-free deposition temperature may result in a prominent ODMR frequency shift near the films. These results may appear counter-intuitive as it is easy to imagine that the diamond's exceptional hardness could resit to the expansion of "soft" thin films. Knowledge of such stresses could help to account for, avoid and even exploit them.

\section{Acknowledgements}

A. Berzins acknowledges support from PostDoc Latvia project 1.1.1.2/VIAA/1/16/024 "Two-way research of thin films and NV centers in diamond crystal" and Latvian Council of Science project lzp-2020/2-0243 "Robust and fast quantum magnetic microscope with concentrated bias field," as well as LLC "MikroTik" donation project, administered by the UoL foundation, "Improvement of Magnetic field imaging system" for the opportunity to significantly improve experimental setup as well as "Simulations for stimulation of science" for the opportunity to acquire COMSOL license. I. Fescenko acknowledges support from ERAF project 1.1.1.5/20/A/001 and LLC "MikroTik" donation project "Annealing furnace for the development of new nanometer-sized sensors and devices," administered by the University of Latvia Foundation. Institute of Solid State Physics, University of Latvia, as the Center of Excellence, has received funding from the European Union's Horizon 2020 Framework Programme H2020-WIDESPREAD-01-2016-2017-TeamingPhase2 under grant agreement No. 739508, project CAMART2. We also acknowledge Laima Busaite for valuable consultations on spin-stress interactions. 

\bibliography{references.bib}

\begin{thebibliography}{56}
\expandafter\ifx\csname natexlab\endcsname\relax\def\natexlab#1{#1}\fi
\expandafter\ifx\csname bibnamefont\endcsname\relax
  \def\bibnamefont#1{#1}\fi
\expandafter\ifx\csname bibfnamefont\endcsname\relax
  \def\bibfnamefont#1{#1}\fi
\expandafter\ifx\csname citenamefont\endcsname\relax
  \def\citenamefont#1{#1}\fi
\expandafter\ifx\csname url\endcsname\relax
  \def\url#1{\texttt{#1}}\fi
\expandafter\ifx\csname urlprefix\endcsname\relax\def\urlprefix{URL }\fi
\providecommand{\bibinfo}[2]{#2}
\providecommand{\eprint}[2][]{\url{#2}}

\bibitem[{\citenamefont{Ashfold et~al.}(2020)\citenamefont{Ashfold, Goss,
  Green, May, Newton, and Peaker}}]{ashfold_nitrogen_2020}
\bibinfo{author}{\bibfnamefont{M.~N.~R.} \bibnamefont{Ashfold}},
  \bibinfo{author}{\bibfnamefont{J.~P.} \bibnamefont{Goss}},
  \bibinfo{author}{\bibfnamefont{B.~L.} \bibnamefont{Green}},
  \bibinfo{author}{\bibfnamefont{P.~W.} \bibnamefont{May}},
  \bibinfo{author}{\bibfnamefont{M.~E.} \bibnamefont{Newton}},
  \bibnamefont{and} \bibinfo{author}{\bibfnamefont{C.~V.}
  \bibnamefont{Peaker}}, \bibinfo{journal}{Chemical Reviews}
  \textbf{\bibinfo{volume}{120}}, \bibinfo{pages}{5745} (\bibinfo{year}{2020}),
  ISSN \bibinfo{issn}{0009-2665, 1520-6890},
  \urlprefix\url{https://pubs.acs.org/doi/10.1021/acs.chemrev.9b00518}.

\bibitem[{\citenamefont{Wu et~al.}(2016)\citenamefont{Wu, Jelezko, Plenio, and
  Weil}}]{wu_diamond_2016}
\bibinfo{author}{\bibfnamefont{Y.}~\bibnamefont{Wu}},
  \bibinfo{author}{\bibfnamefont{F.}~\bibnamefont{Jelezko}},
  \bibinfo{author}{\bibfnamefont{M.~B.} \bibnamefont{Plenio}},
  \bibnamefont{and} \bibinfo{author}{\bibfnamefont{T.}~\bibnamefont{Weil}},
  \bibinfo{journal}{Angewandte Chemie International Edition}
  \textbf{\bibinfo{volume}{55}}, \bibinfo{pages}{6586} (\bibinfo{year}{2016}),
  ISSN \bibinfo{issn}{14337851},
  \urlprefix\url{http://doi.wiley.com/10.1002/anie.201506556}.

\bibitem[{\citenamefont{Chipaux et~al.}(2018)\citenamefont{Chipaux, van~der
  Laan, Hemelaar, Hasani, Zheng, and Schirhagl}}]{chipaux_nanodiamonds_2018}
\bibinfo{author}{\bibfnamefont{M.}~\bibnamefont{Chipaux}},
  \bibinfo{author}{\bibfnamefont{K.~J.} \bibnamefont{van~der Laan}},
  \bibinfo{author}{\bibfnamefont{S.~R.} \bibnamefont{Hemelaar}},
  \bibinfo{author}{\bibfnamefont{M.}~\bibnamefont{Hasani}},
  \bibinfo{author}{\bibfnamefont{T.}~\bibnamefont{Zheng}}, \bibnamefont{and}
  \bibinfo{author}{\bibfnamefont{R.}~\bibnamefont{Schirhagl}},
  \bibinfo{journal}{Small} \textbf{\bibinfo{volume}{14}},
  \bibinfo{pages}{1704263} (\bibinfo{year}{2018}), ISSN
  \bibinfo{issn}{16136810},
  \urlprefix\url{http://doi.wiley.com/10.1002/smll.201704263}.

\bibitem[{\citenamefont{Norman et~al.}(2020)\citenamefont{Norman, Majety, Wang,
  Casey, Curro, and Radulaski}}]{norman_novel_2020}
\bibinfo{author}{\bibfnamefont{V.~A.} \bibnamefont{Norman}},
  \bibinfo{author}{\bibfnamefont{S.}~\bibnamefont{Majety}},
  \bibinfo{author}{\bibfnamefont{Z.}~\bibnamefont{Wang}},
  \bibinfo{author}{\bibfnamefont{W.~H.} \bibnamefont{Casey}},
  \bibinfo{author}{\bibfnamefont{N.}~\bibnamefont{Curro}}, \bibnamefont{and}
  \bibinfo{author}{\bibfnamefont{M.}~\bibnamefont{Radulaski}},
  \bibinfo{journal}{InfoMat} p. \bibinfo{pages}{inf2.12128}
  (\bibinfo{year}{2020}), ISSN \bibinfo{issn}{2567-3165, 2567-3165},
  \urlprefix\url{https://onlinelibrary.wiley.com/doi/abs/10.1002/inf2.12128}.

\bibitem[{\citenamefont{Fu et~al.}(2020)\citenamefont{Fu, Iwata, Wickenbrock,
  and Budker}}]{fu_sensitive_2020}
\bibinfo{author}{\bibfnamefont{K.-M.~C.} \bibnamefont{Fu}},
  \bibinfo{author}{\bibfnamefont{G.~Z.} \bibnamefont{Iwata}},
  \bibinfo{author}{\bibfnamefont{A.}~\bibnamefont{Wickenbrock}},
  \bibnamefont{and} \bibinfo{author}{\bibfnamefont{D.}~\bibnamefont{Budker}},
  \bibinfo{journal}{AVS Quantum Science} \textbf{\bibinfo{volume}{2}},
  \bibinfo{pages}{044702} (\bibinfo{year}{2020}), ISSN
  \bibinfo{issn}{2639-0213},
  \urlprefix\url{http://avs.scitation.org/doi/10.1116/5.0025186}.

\bibitem[{\citenamefont{Abe and Sasaki}(2018)}]{abe_tutorial_2018}
\bibinfo{author}{\bibfnamefont{E.}~\bibnamefont{Abe}} \bibnamefont{and}
  \bibinfo{author}{\bibfnamefont{K.}~\bibnamefont{Sasaki}},
  \bibinfo{journal}{Journal of Applied Physics} \textbf{\bibinfo{volume}{123}},
  \bibinfo{pages}{161101} (\bibinfo{year}{2018}), ISSN
  \bibinfo{issn}{0021-8979, 1089-7550},
  \urlprefix\url{http://aip.scitation.org/doi/10.1063/1.5011231}.

\bibitem[{\citenamefont{Rondin et~al.}(2014)\citenamefont{Rondin, Tetienne,
  Hingant, Roch, Maletinsky, and Jacques}}]{rondin_magnetometry_2014}
\bibinfo{author}{\bibfnamefont{L.}~\bibnamefont{Rondin}},
  \bibinfo{author}{\bibfnamefont{J.-P.} \bibnamefont{Tetienne}},
  \bibinfo{author}{\bibfnamefont{T.}~\bibnamefont{Hingant}},
  \bibinfo{author}{\bibfnamefont{J.-F.} \bibnamefont{Roch}},
  \bibinfo{author}{\bibfnamefont{P.}~\bibnamefont{Maletinsky}},
  \bibnamefont{and} \bibinfo{author}{\bibfnamefont{V.}~\bibnamefont{Jacques}},
  \bibinfo{journal}{Reports on Progress in Physics}
  \textbf{\bibinfo{volume}{77}}, \bibinfo{pages}{056503}
  (\bibinfo{year}{2014}), ISSN \bibinfo{issn}{0034-4885, 1361-6633},
  \urlprefix\url{https://iopscience.iop.org/article/10.1088/0034-4885/77/5/056503}.

\bibitem[{\citenamefont{Aharonovich et~al.}(2011)\citenamefont{Aharonovich,
  Greentree, and Prawer}}]{aharonovich_diamond_2011}
\bibinfo{author}{\bibfnamefont{I.}~\bibnamefont{Aharonovich}},
  \bibinfo{author}{\bibfnamefont{A.~D.} \bibnamefont{Greentree}},
  \bibnamefont{and} \bibinfo{author}{\bibfnamefont{S.}~\bibnamefont{Prawer}},
  \bibinfo{journal}{Nature Photonics} \textbf{\bibinfo{volume}{5}},
  \bibinfo{pages}{397} (\bibinfo{year}{2011}), ISSN \bibinfo{issn}{1749-4885,
  1749-4893}, \urlprefix\url{http://www.nature.com/articles/nphoton.2011.54}.

\bibitem[{\citenamefont{Schukraft et~al.}(2016)\citenamefont{Schukraft, Zheng,
  Schröder, Mouradian, Walsh, Trusheim, Bakhru, and
  Englund}}]{schukraft_invited_2016}
\bibinfo{author}{\bibfnamefont{M.}~\bibnamefont{Schukraft}},
  \bibinfo{author}{\bibfnamefont{J.}~\bibnamefont{Zheng}},
  \bibinfo{author}{\bibfnamefont{T.}~\bibnamefont{Schröder}},
  \bibinfo{author}{\bibfnamefont{S.~L.} \bibnamefont{Mouradian}},
  \bibinfo{author}{\bibfnamefont{M.}~\bibnamefont{Walsh}},
  \bibinfo{author}{\bibfnamefont{M.~E.} \bibnamefont{Trusheim}},
  \bibinfo{author}{\bibfnamefont{H.}~\bibnamefont{Bakhru}}, \bibnamefont{and}
  \bibinfo{author}{\bibfnamefont{D.~R.} \bibnamefont{Englund}},
  \bibinfo{journal}{APL Photonics} \textbf{\bibinfo{volume}{1}},
  \bibinfo{pages}{020801} (\bibinfo{year}{2016}), ISSN
  \bibinfo{issn}{2378-0967},
  \urlprefix\url{http://aip.scitation.org/doi/10.1063/1.4948746}.

\bibitem[{\citenamefont{Ruf et~al.}(2019)\citenamefont{Ruf, IJspeert, van Dam,
  de~Jong, van~den Berg, Evers, and Hanson}}]{ruf_optically_2019}
\bibinfo{author}{\bibfnamefont{M.}~\bibnamefont{Ruf}},
  \bibinfo{author}{\bibfnamefont{M.}~\bibnamefont{IJspeert}},
  \bibinfo{author}{\bibfnamefont{S.}~\bibnamefont{van Dam}},
  \bibinfo{author}{\bibfnamefont{N.}~\bibnamefont{de~Jong}},
  \bibinfo{author}{\bibfnamefont{H.}~\bibnamefont{van~den Berg}},
  \bibinfo{author}{\bibfnamefont{G.}~\bibnamefont{Evers}}, \bibnamefont{and}
  \bibinfo{author}{\bibfnamefont{R.}~\bibnamefont{Hanson}},
  \bibinfo{journal}{Nano Letters} \textbf{\bibinfo{volume}{19}},
  \bibinfo{pages}{3987} (\bibinfo{year}{2019}), ISSN \bibinfo{issn}{1530-6984,
  1530-6992},
  \urlprefix\url{https://pubs.acs.org/doi/10.1021/acs.nanolett.9b01316}.

\bibitem[{\citenamefont{Golter et~al.}(2016)\citenamefont{Golter, Oo, Amezcua,
  Stewart, and Wang}}]{golter_optomechanical_2016}
\bibinfo{author}{\bibfnamefont{D.~A.} \bibnamefont{Golter}},
  \bibinfo{author}{\bibfnamefont{T.}~\bibnamefont{Oo}},
  \bibinfo{author}{\bibfnamefont{M.}~\bibnamefont{Amezcua}},
  \bibinfo{author}{\bibfnamefont{K.~A.} \bibnamefont{Stewart}},
  \bibnamefont{and} \bibinfo{author}{\bibfnamefont{H.}~\bibnamefont{Wang}},
  \bibinfo{journal}{Physical Review Letters} \textbf{\bibinfo{volume}{116}},
  \bibinfo{pages}{143602} (\bibinfo{year}{2016}), ISSN
  \bibinfo{issn}{0031-9007, 1079-7114},
  \urlprefix\url{https://link.aps.org/doi/10.1103/PhysRevLett.116.143602}.

\bibitem[{\citenamefont{Labanowski et~al.}(2018)\citenamefont{Labanowski,
  Bhallamudi, Guo, Purser, McCullian, Hammel, and
  Salahuddin}}]{labanowski_voltage-driven_2018}
\bibinfo{author}{\bibfnamefont{D.}~\bibnamefont{Labanowski}},
  \bibinfo{author}{\bibfnamefont{V.~P.} \bibnamefont{Bhallamudi}},
  \bibinfo{author}{\bibfnamefont{Q.}~\bibnamefont{Guo}},
  \bibinfo{author}{\bibfnamefont{C.~M.} \bibnamefont{Purser}},
  \bibinfo{author}{\bibfnamefont{B.~A.} \bibnamefont{McCullian}},
  \bibinfo{author}{\bibfnamefont{P.~C.} \bibnamefont{Hammel}},
  \bibnamefont{and}
  \bibinfo{author}{\bibfnamefont{S.}~\bibnamefont{Salahuddin}},
  \bibinfo{journal}{Science Advances} \textbf{\bibinfo{volume}{4}},
  \bibinfo{pages}{eaat6574} (\bibinfo{year}{2018}), ISSN
  \bibinfo{issn}{2375-2548},
  \urlprefix\url{https://www.science.org/doi/10.1126/sciadv.aat6574}.

\bibitem[{\citenamefont{Xu et~al.}(2019)\citenamefont{Xu, Yin, Han, and
  Li}}]{xu_quantum_2019}
\bibinfo{author}{\bibfnamefont{Z.}~\bibnamefont{Xu}},
  \bibinfo{author}{\bibfnamefont{Z.-q.} \bibnamefont{Yin}},
  \bibinfo{author}{\bibfnamefont{Q.}~\bibnamefont{Han}}, \bibnamefont{and}
  \bibinfo{author}{\bibfnamefont{T.}~\bibnamefont{Li}},
  \bibinfo{journal}{Optical Materials Express} \textbf{\bibinfo{volume}{9}},
  \bibinfo{pages}{4654} (\bibinfo{year}{2019}), ISSN \bibinfo{issn}{2159-3930},
  \urlprefix\url{https://www.osapublishing.org/abstract.cfm?URI=ome-9-12-4654}.

\bibitem[{\citenamefont{Bersin et~al.}(2019)\citenamefont{Bersin, Walsh,
  Mouradian, Trusheim, Schröder, and Englund}}]{bersin_individual_2019}
\bibinfo{author}{\bibfnamefont{E.}~\bibnamefont{Bersin}},
  \bibinfo{author}{\bibfnamefont{M.}~\bibnamefont{Walsh}},
  \bibinfo{author}{\bibfnamefont{S.~L.} \bibnamefont{Mouradian}},
  \bibinfo{author}{\bibfnamefont{M.~E.} \bibnamefont{Trusheim}},
  \bibinfo{author}{\bibfnamefont{T.}~\bibnamefont{Schröder}},
  \bibnamefont{and} \bibinfo{author}{\bibfnamefont{D.}~\bibnamefont{Englund}},
  \bibinfo{journal}{npj Quantum Information} \textbf{\bibinfo{volume}{5}},
  \bibinfo{pages}{38} (\bibinfo{year}{2019}), ISSN \bibinfo{issn}{2056-6387},
  \urlprefix\url{http://www.nature.com/articles/s41534-019-0154-y}.

\bibitem[{\citenamefont{Udvarhelyi et~al.}(2018)\citenamefont{Udvarhelyi,
  Shkolnikov, Gali, Burkard, and Pályi}}]{udvarhelyi_spin-strain_2018}
\bibinfo{author}{\bibfnamefont{P.}~\bibnamefont{Udvarhelyi}},
  \bibinfo{author}{\bibfnamefont{V.~O.} \bibnamefont{Shkolnikov}},
  \bibinfo{author}{\bibfnamefont{A.}~\bibnamefont{Gali}},
  \bibinfo{author}{\bibfnamefont{G.}~\bibnamefont{Burkard}}, \bibnamefont{and}
  \bibinfo{author}{\bibfnamefont{A.}~\bibnamefont{Pályi}},
  \bibinfo{journal}{Physical Review B} \textbf{\bibinfo{volume}{98}},
  \bibinfo{pages}{075201} (\bibinfo{year}{2018}), ISSN
  \bibinfo{issn}{2469-9950, 2469-9969},
  \urlprefix\url{https://link.aps.org/doi/10.1103/PhysRevB.98.075201}.

\bibitem[{\citenamefont{Li et~al.}(2014)\citenamefont{Li, Shan, and
  Ma}}]{li_elastic_2014}
\bibinfo{author}{\bibfnamefont{J.}~\bibnamefont{Li}},
  \bibinfo{author}{\bibfnamefont{Z.}~\bibnamefont{Shan}}, \bibnamefont{and}
  \bibinfo{author}{\bibfnamefont{E.}~\bibnamefont{Ma}}, \bibinfo{journal}{MRS
  Bulletin} \textbf{\bibinfo{volume}{39}}, \bibinfo{pages}{108}
  (\bibinfo{year}{2014}), ISSN \bibinfo{issn}{0883-7694, 1938-1425},
  \urlprefix\url{http://link.springer.com/10.1557/mrs.2014.3}.

\bibitem[{\citenamefont{Smith et~al.}(2009)\citenamefont{Smith, Mohs, and
  Nie}}]{smith_tuning_2009}
\bibinfo{author}{\bibfnamefont{A.~M.} \bibnamefont{Smith}},
  \bibinfo{author}{\bibfnamefont{A.~M.} \bibnamefont{Mohs}}, \bibnamefont{and}
  \bibinfo{author}{\bibfnamefont{S.}~\bibnamefont{Nie}},
  \bibinfo{journal}{Nature Nanotechnology} \textbf{\bibinfo{volume}{4}},
  \bibinfo{pages}{56} (\bibinfo{year}{2009}), ISSN \bibinfo{issn}{1748-3387,
  1748-3395}, \urlprefix\url{http://www.nature.com/articles/nnano.2008.360}.

\bibitem[{\citenamefont{Barson et~al.}(2017)\citenamefont{Barson, Peddibhotla,
  Ovartchaiyapong, Ganesan, Taylor, Gebert, Mielens, Koslowski, Simpson,
  McGuinness et~al.}}]{barson_nanomechanical_2017}
\bibinfo{author}{\bibfnamefont{M.~S.~J.} \bibnamefont{Barson}},
  \bibinfo{author}{\bibfnamefont{P.}~\bibnamefont{Peddibhotla}},
  \bibinfo{author}{\bibfnamefont{P.}~\bibnamefont{Ovartchaiyapong}},
  \bibinfo{author}{\bibfnamefont{K.}~\bibnamefont{Ganesan}},
  \bibinfo{author}{\bibfnamefont{R.~L.} \bibnamefont{Taylor}},
  \bibinfo{author}{\bibfnamefont{M.}~\bibnamefont{Gebert}},
  \bibinfo{author}{\bibfnamefont{Z.}~\bibnamefont{Mielens}},
  \bibinfo{author}{\bibfnamefont{B.}~\bibnamefont{Koslowski}},
  \bibinfo{author}{\bibfnamefont{D.~A.} \bibnamefont{Simpson}},
  \bibinfo{author}{\bibfnamefont{L.~P.} \bibnamefont{McGuinness}},
  \bibnamefont{et~al.}, \bibinfo{journal}{Nano Letters}
  \textbf{\bibinfo{volume}{17}}, \bibinfo{pages}{1496} (\bibinfo{year}{2017}),
  ISSN \bibinfo{issn}{1530-6984, 1530-6992},
  \urlprefix\url{https://pubs.acs.org/doi/10.1021/acs.nanolett.6b04544}.

\bibitem[{\citenamefont{Barfuss et~al.}(2019)\citenamefont{Barfuss, Kasperczyk,
  Kölbl, and Maletinsky}}]{barfuss_spin-stress_2019}
\bibinfo{author}{\bibfnamefont{A.}~\bibnamefont{Barfuss}},
  \bibinfo{author}{\bibfnamefont{M.}~\bibnamefont{Kasperczyk}},
  \bibinfo{author}{\bibfnamefont{J.}~\bibnamefont{Kölbl}}, \bibnamefont{and}
  \bibinfo{author}{\bibfnamefont{P.}~\bibnamefont{Maletinsky}},
  \bibinfo{journal}{Physical Review B} \textbf{\bibinfo{volume}{99}},
  \bibinfo{pages}{174102} (\bibinfo{year}{2019}), ISSN
  \bibinfo{issn}{2469-9950, 2469-9969},
  \urlprefix\url{https://link.aps.org/doi/10.1103/PhysRevB.99.174102}.

\bibitem[{\citenamefont{Kehayias et~al.}(2019)\citenamefont{Kehayias, Turner,
  Trubko, Schloss, Hart, Wesson, Glenn, and Walsworth}}]{kehayias_imaging_2019}
\bibinfo{author}{\bibfnamefont{P.}~\bibnamefont{Kehayias}},
  \bibinfo{author}{\bibfnamefont{M.~J.} \bibnamefont{Turner}},
  \bibinfo{author}{\bibfnamefont{R.}~\bibnamefont{Trubko}},
  \bibinfo{author}{\bibfnamefont{J.~M.} \bibnamefont{Schloss}},
  \bibinfo{author}{\bibfnamefont{C.~A.} \bibnamefont{Hart}},
  \bibinfo{author}{\bibfnamefont{M.}~\bibnamefont{Wesson}},
  \bibinfo{author}{\bibfnamefont{D.~R.} \bibnamefont{Glenn}}, \bibnamefont{and}
  \bibinfo{author}{\bibfnamefont{R.~L.} \bibnamefont{Walsworth}},
  \bibinfo{journal}{Physical Review B} \textbf{\bibinfo{volume}{100}},
  \bibinfo{pages}{174103} (\bibinfo{year}{2019}), \bibinfo{note}{publisher:
  American Physical Society},
  \urlprefix\url{https://link.aps.org/doi/10.1103/PhysRevB.100.174103}.

\bibitem[{\citenamefont{Broadway et~al.}(2019)\citenamefont{Broadway, Johnson,
  Barson, Lillie, Dontschuk, McCloskey, Tsai, Teraji, Simpson, Stacey
  et~al.}}]{broadway_microscopic_2019}
\bibinfo{author}{\bibfnamefont{D.~A.} \bibnamefont{Broadway}},
  \bibinfo{author}{\bibfnamefont{B.~C.} \bibnamefont{Johnson}},
  \bibinfo{author}{\bibfnamefont{M.~S.~J.} \bibnamefont{Barson}},
  \bibinfo{author}{\bibfnamefont{S.~E.} \bibnamefont{Lillie}},
  \bibinfo{author}{\bibfnamefont{N.}~\bibnamefont{Dontschuk}},
  \bibinfo{author}{\bibfnamefont{D.~J.} \bibnamefont{McCloskey}},
  \bibinfo{author}{\bibfnamefont{A.}~\bibnamefont{Tsai}},
  \bibinfo{author}{\bibfnamefont{T.}~\bibnamefont{Teraji}},
  \bibinfo{author}{\bibfnamefont{D.~A.} \bibnamefont{Simpson}},
  \bibinfo{author}{\bibfnamefont{A.}~\bibnamefont{Stacey}},
  \bibnamefont{et~al.}, \bibinfo{journal}{Nano Letters}
  \textbf{\bibinfo{volume}{19}}, \bibinfo{pages}{4543} (\bibinfo{year}{2019}),
  ISSN \bibinfo{issn}{1530-6984, 1530-6992},
  \urlprefix\url{https://pubs.acs.org/doi/10.1021/acs.nanolett.9b01402}.

\bibitem[{\citenamefont{Trusheim and Englund}(2016)}]{trusheim_wide-field_2016}
\bibinfo{author}{\bibfnamefont{M.~E.} \bibnamefont{Trusheim}} \bibnamefont{and}
  \bibinfo{author}{\bibfnamefont{D.}~\bibnamefont{Englund}},
  \bibinfo{journal}{New Journal of Physics} \textbf{\bibinfo{volume}{18}},
  \bibinfo{pages}{123023} (\bibinfo{year}{2016}), ISSN
  \bibinfo{issn}{1367-2630}, \bibinfo{note}{publisher: IOP Publishing},
  \urlprefix\url{https://doi.org/10.1088%2F1367-2630%2Faa5040}.

\bibitem[{\citenamefont{Ho et~al.}(2021)\citenamefont{Ho, Leung, Pang, Wong,
  Ng, and Yang}}]{ho__2021}
\bibinfo{author}{\bibfnamefont{K.~O.} \bibnamefont{Ho}},
  \bibinfo{author}{\bibfnamefont{M.~Y.} \bibnamefont{Leung}},
  \bibinfo{author}{\bibfnamefont{Y.~Y.} \bibnamefont{Pang}},
  \bibinfo{author}{\bibfnamefont{K.~C.} \bibnamefont{Wong}},
  \bibinfo{author}{\bibfnamefont{P.~H.} \bibnamefont{Ng}}, \bibnamefont{and}
  \bibinfo{author}{\bibfnamefont{S.}~\bibnamefont{Yang}}, \bibinfo{journal}{ACS
  Applied Polymer Materials} \textbf{\bibinfo{volume}{3}}, \bibinfo{pages}{162}
  (\bibinfo{year}{2021}), ISSN \bibinfo{issn}{2637-6105, 2637-6105},
  \urlprefix\url{https://pubs.acs.org/doi/10.1021/acsapm.0c00964}.

\bibitem[{\citenamefont{Ho et~al.}(2020)\citenamefont{Ho, Leung, Jiang, Ao,
  Zhang, Yip, Pang, Wong, Goh, and Yang}}]{ho_probing_2020}
\bibinfo{author}{\bibfnamefont{K.~O.} \bibnamefont{Ho}},
  \bibinfo{author}{\bibfnamefont{M.~Y.} \bibnamefont{Leung}},
  \bibinfo{author}{\bibfnamefont{Y.}~\bibnamefont{Jiang}},
  \bibinfo{author}{\bibfnamefont{K.~P.} \bibnamefont{Ao}},
  \bibinfo{author}{\bibfnamefont{W.}~\bibnamefont{Zhang}},
  \bibinfo{author}{\bibfnamefont{K.~Y.} \bibnamefont{Yip}},
  \bibinfo{author}{\bibfnamefont{Y.~Y.} \bibnamefont{Pang}},
  \bibinfo{author}{\bibfnamefont{K.~C.} \bibnamefont{Wong}},
  \bibinfo{author}{\bibfnamefont{S.~K.} \bibnamefont{Goh}}, \bibnamefont{and}
  \bibinfo{author}{\bibfnamefont{S.}~\bibnamefont{Yang}},
  \bibinfo{journal}{Physical Review Applied} \textbf{\bibinfo{volume}{13}},
  \bibinfo{pages}{024041} (\bibinfo{year}{2020}), ISSN
  \bibinfo{issn}{2331-7019},
  \urlprefix\url{https://link.aps.org/doi/10.1103/PhysRevApplied.13.024041}.

\bibitem[{\citenamefont{Hsieh et~al.}(2019)\citenamefont{Hsieh, Bhattacharyya,
  Zu, Mittiga, Smart, Machado, Kobrin, Höhn, Rui, Kamrani
  et~al.}}]{hsieh_imaging_2019}
\bibinfo{author}{\bibfnamefont{S.}~\bibnamefont{Hsieh}},
  \bibinfo{author}{\bibfnamefont{P.}~\bibnamefont{Bhattacharyya}},
  \bibinfo{author}{\bibfnamefont{C.}~\bibnamefont{Zu}},
  \bibinfo{author}{\bibfnamefont{T.}~\bibnamefont{Mittiga}},
  \bibinfo{author}{\bibfnamefont{T.~J.} \bibnamefont{Smart}},
  \bibinfo{author}{\bibfnamefont{F.}~\bibnamefont{Machado}},
  \bibinfo{author}{\bibfnamefont{B.}~\bibnamefont{Kobrin}},
  \bibinfo{author}{\bibfnamefont{T.~O.} \bibnamefont{Höhn}},
  \bibinfo{author}{\bibfnamefont{N.~Z.} \bibnamefont{Rui}},
  \bibinfo{author}{\bibfnamefont{M.}~\bibnamefont{Kamrani}},
  \bibnamefont{et~al.}, \bibinfo{journal}{Science}
  \textbf{\bibinfo{volume}{366}}, \bibinfo{pages}{1349} (\bibinfo{year}{2019}),
  ISSN \bibinfo{issn}{0036-8075, 1095-9203},
  \urlprefix\url{https://www.science.org/doi/10.1126/science.aaw4352}.

\bibitem[{\citenamefont{Marshall
  et~al.}(2021{\natexlab{a}})\citenamefont{Marshall, Turner, Ku, Phillips, and
  Walsworth}}]{marshall_directional_2021}
\bibinfo{author}{\bibfnamefont{M.~C.} \bibnamefont{Marshall}},
  \bibinfo{author}{\bibfnamefont{M.~J.} \bibnamefont{Turner}},
  \bibinfo{author}{\bibfnamefont{M.~J.~H.} \bibnamefont{Ku}},
  \bibinfo{author}{\bibfnamefont{D.~F.} \bibnamefont{Phillips}},
  \bibnamefont{and} \bibinfo{author}{\bibfnamefont{R.~L.}
  \bibnamefont{Walsworth}}, \bibinfo{journal}{Quantum Science and Technology}
  \textbf{\bibinfo{volume}{6}}, \bibinfo{pages}{024011}
  (\bibinfo{year}{2021}{\natexlab{a}}), ISSN \bibinfo{issn}{2058-9565},
  \urlprefix\url{https://iopscience.iop.org/article/10.1088/2058-9565/abe5ed}.

\bibitem[{\citenamefont{Rajendran et~al.}(2017)\citenamefont{Rajendran,
  Zobrist, Sushkov, Walsworth, and Lukin}}]{rajendran_method_2017}
\bibinfo{author}{\bibfnamefont{S.}~\bibnamefont{Rajendran}},
  \bibinfo{author}{\bibfnamefont{N.}~\bibnamefont{Zobrist}},
  \bibinfo{author}{\bibfnamefont{A.~O.} \bibnamefont{Sushkov}},
  \bibinfo{author}{\bibfnamefont{R.}~\bibnamefont{Walsworth}},
  \bibnamefont{and} \bibinfo{author}{\bibfnamefont{M.}~\bibnamefont{Lukin}},
  \bibinfo{journal}{Physical Review D} \textbf{\bibinfo{volume}{96}},
  \bibinfo{pages}{035009} (\bibinfo{year}{2017}), ISSN
  \bibinfo{issn}{2470-0010, 2470-0029},
  \urlprefix\url{https://link.aps.org/doi/10.1103/PhysRevD.96.035009}.

\bibitem[{\citenamefont{Teissier et~al.}(2014)\citenamefont{Teissier, Barfuss,
  Appel, Neu, and Maletinsky}}]{teissier_strain_2014}
\bibinfo{author}{\bibfnamefont{J.}~\bibnamefont{Teissier}},
  \bibinfo{author}{\bibfnamefont{A.}~\bibnamefont{Barfuss}},
  \bibinfo{author}{\bibfnamefont{P.}~\bibnamefont{Appel}},
  \bibinfo{author}{\bibfnamefont{E.}~\bibnamefont{Neu}}, \bibnamefont{and}
  \bibinfo{author}{\bibfnamefont{P.}~\bibnamefont{Maletinsky}},
  \bibinfo{journal}{Physical Review Letters} \textbf{\bibinfo{volume}{113}},
  \bibinfo{pages}{020503} (\bibinfo{year}{2014}), ISSN
  \bibinfo{issn}{0031-9007, 1079-7114},
  \urlprefix\url{https://link.aps.org/doi/10.1103/PhysRevLett.113.020503}.

\bibitem[{\citenamefont{Acosta et~al.}(2010)\citenamefont{Acosta, Bauch,
  Ledbetter, Waxman, Bouchard, and Budker}}]{acosta_temperature_2010}
\bibinfo{author}{\bibfnamefont{V.~M.} \bibnamefont{Acosta}},
  \bibinfo{author}{\bibfnamefont{E.}~\bibnamefont{Bauch}},
  \bibinfo{author}{\bibfnamefont{M.~P.} \bibnamefont{Ledbetter}},
  \bibinfo{author}{\bibfnamefont{A.}~\bibnamefont{Waxman}},
  \bibinfo{author}{\bibfnamefont{L.-S.} \bibnamefont{Bouchard}},
  \bibnamefont{and} \bibinfo{author}{\bibfnamefont{D.}~\bibnamefont{Budker}},
  \bibinfo{journal}{Physical Review Letters} \textbf{\bibinfo{volume}{104}},
  \bibinfo{pages}{070801} (\bibinfo{year}{2010}), ISSN
  \bibinfo{issn}{0031-9007, 1079-7114},
  \urlprefix\url{https://link.aps.org/doi/10.1103/PhysRevLett.104.070801}.

\bibitem[{\citenamefont{Doherty et~al.}(2012)\citenamefont{Doherty, Dolde,
  Fedder, Jelezko, Wrachtrup, Manson, and Hollenberg}}]{doherty_theory_2012}
\bibinfo{author}{\bibfnamefont{M.~W.} \bibnamefont{Doherty}},
  \bibinfo{author}{\bibfnamefont{F.}~\bibnamefont{Dolde}},
  \bibinfo{author}{\bibfnamefont{H.}~\bibnamefont{Fedder}},
  \bibinfo{author}{\bibfnamefont{F.}~\bibnamefont{Jelezko}},
  \bibinfo{author}{\bibfnamefont{J.}~\bibnamefont{Wrachtrup}},
  \bibinfo{author}{\bibfnamefont{N.~B.} \bibnamefont{Manson}},
  \bibnamefont{and} \bibinfo{author}{\bibfnamefont{L.~C.~L.}
  \bibnamefont{Hollenberg}}, \bibinfo{journal}{Physical Review B}
  \textbf{\bibinfo{volume}{85}}, \bibinfo{pages}{205203}
  (\bibinfo{year}{2012}), \bibinfo{note}{publisher: American Physical Society},
  \urlprefix\url{https://link.aps.org/doi/10.1103/PhysRevB.85.205203}.

\bibitem[{\citenamefont{Knauer et~al.}(2020)\citenamefont{Knauer, Hadden, and
  Rarity}}]{knauer_-situ_2020}
\bibinfo{author}{\bibfnamefont{S.}~\bibnamefont{Knauer}},
  \bibinfo{author}{\bibfnamefont{J.~P.} \bibnamefont{Hadden}},
  \bibnamefont{and} \bibinfo{author}{\bibfnamefont{J.~G.}
  \bibnamefont{Rarity}}, \bibinfo{journal}{npj Quantum Information}
  \textbf{\bibinfo{volume}{6}}, \bibinfo{pages}{1} (\bibinfo{year}{2020}), ISSN
  \bibinfo{issn}{2056-6387}, \bibinfo{note}{number: 1 Publisher: Nature
  Publishing Group},
  \urlprefix\url{https://www.nature.com/articles/s41534-020-0277-1}.

\bibitem[{\citenamefont{Wood et~al.}(2016)\citenamefont{Wood, Broadway, Hall,
  Stacey, Simpson, Tetienne, and Hollenberg}}]{wood_wide-band_2016}
\bibinfo{author}{\bibfnamefont{J.~D.~A.} \bibnamefont{Wood}},
  \bibinfo{author}{\bibfnamefont{D.~A.} \bibnamefont{Broadway}},
  \bibinfo{author}{\bibfnamefont{L.~T.} \bibnamefont{Hall}},
  \bibinfo{author}{\bibfnamefont{A.}~\bibnamefont{Stacey}},
  \bibinfo{author}{\bibfnamefont{D.~A.} \bibnamefont{Simpson}},
  \bibinfo{author}{\bibfnamefont{J.-P.} \bibnamefont{Tetienne}},
  \bibnamefont{and} \bibinfo{author}{\bibfnamefont{L.~C.~L.}
  \bibnamefont{Hollenberg}}, \bibinfo{journal}{Physical Review B}
  \textbf{\bibinfo{volume}{94}}, \bibinfo{pages}{155402}
  (\bibinfo{year}{2016}), ISSN \bibinfo{issn}{2469-9950, 2469-9969},
  \urlprefix\url{https://link.aps.org/doi/10.1103/PhysRevB.94.155402}.

\bibitem[{\citenamefont{Lazda et~al.}(2021)\citenamefont{Lazda, Busaite,
  Berzins, Smits, Gahbauer, Auzinsh, Budker, and
  Ferber}}]{lazda_cross-relaxation_2021}
\bibinfo{author}{\bibfnamefont{R.}~\bibnamefont{Lazda}},
  \bibinfo{author}{\bibfnamefont{L.}~\bibnamefont{Busaite}},
  \bibinfo{author}{\bibfnamefont{A.}~\bibnamefont{Berzins}},
  \bibinfo{author}{\bibfnamefont{J.}~\bibnamefont{Smits}},
  \bibinfo{author}{\bibfnamefont{F.}~\bibnamefont{Gahbauer}},
  \bibinfo{author}{\bibfnamefont{M.}~\bibnamefont{Auzinsh}},
  \bibinfo{author}{\bibfnamefont{D.}~\bibnamefont{Budker}}, \bibnamefont{and}
  \bibinfo{author}{\bibfnamefont{R.}~\bibnamefont{Ferber}},
  \bibinfo{journal}{Physical Review B} \textbf{\bibinfo{volume}{103}},
  \bibinfo{pages}{134104} (\bibinfo{year}{2021}), ISSN
  \bibinfo{issn}{2469-9950, 2469-9969},
  \urlprefix\url{https://link.aps.org/doi/10.1103/PhysRevB.103.134104}.

\bibitem[{\citenamefont{Busaite et~al.}(2020)\citenamefont{Busaite, Lazda,
  Berzins, Auzinsh, Ferber, and Gahbauer}}]{busaite_dynamic_2020}
\bibinfo{author}{\bibfnamefont{L.}~\bibnamefont{Busaite}},
  \bibinfo{author}{\bibfnamefont{R.}~\bibnamefont{Lazda}},
  \bibinfo{author}{\bibfnamefont{A.}~\bibnamefont{Berzins}},
  \bibinfo{author}{\bibfnamefont{M.}~\bibnamefont{Auzinsh}},
  \bibinfo{author}{\bibfnamefont{R.}~\bibnamefont{Ferber}}, \bibnamefont{and}
  \bibinfo{author}{\bibfnamefont{F.}~\bibnamefont{Gahbauer}},
  \bibinfo{journal}{Physical Review B} \textbf{\bibinfo{volume}{102}},
  \bibinfo{pages}{224101} (\bibinfo{year}{2020}), ISSN
  \bibinfo{issn}{2469-9950, 2469-9969},
  \urlprefix\url{https://link.aps.org/doi/10.1103/PhysRevB.102.224101}.

\bibitem[{\citenamefont{Fescenko et~al.}(2019)\citenamefont{Fescenko, Laraoui,
  Smits, Mosavian, Kehayias, Seto, Bougas, Jarmola, and
  Acosta}}]{fescenko_diamond_2019}
\bibinfo{author}{\bibfnamefont{I.}~\bibnamefont{Fescenko}},
  \bibinfo{author}{\bibfnamefont{A.}~\bibnamefont{Laraoui}},
  \bibinfo{author}{\bibfnamefont{J.}~\bibnamefont{Smits}},
  \bibinfo{author}{\bibfnamefont{N.}~\bibnamefont{Mosavian}},
  \bibinfo{author}{\bibfnamefont{P.}~\bibnamefont{Kehayias}},
  \bibinfo{author}{\bibfnamefont{J.}~\bibnamefont{Seto}},
  \bibinfo{author}{\bibfnamefont{L.}~\bibnamefont{Bougas}},
  \bibinfo{author}{\bibfnamefont{A.}~\bibnamefont{Jarmola}}, \bibnamefont{and}
  \bibinfo{author}{\bibfnamefont{V.~M.} \bibnamefont{Acosta}},
  \bibinfo{journal}{Physical Review Applied} \textbf{\bibinfo{volume}{11}},
  \bibinfo{pages}{034029} (\bibinfo{year}{2019}), \bibinfo{note}{publisher:
  American Physical Society},
  \urlprefix\url{https://link.aps.org/doi/10.1103/PhysRevApplied.11.034029}.

\bibitem[{\citenamefont{Fescenko et~al.}(2020)\citenamefont{Fescenko, Jarmola,
  Savukov, Kehayias, Smits, Damron, Ristoff, Mosavian, and
  Acosta}}]{fescenko_diamond_2020}
\bibinfo{author}{\bibfnamefont{I.}~\bibnamefont{Fescenko}},
  \bibinfo{author}{\bibfnamefont{A.}~\bibnamefont{Jarmola}},
  \bibinfo{author}{\bibfnamefont{I.}~\bibnamefont{Savukov}},
  \bibinfo{author}{\bibfnamefont{P.}~\bibnamefont{Kehayias}},
  \bibinfo{author}{\bibfnamefont{J.}~\bibnamefont{Smits}},
  \bibinfo{author}{\bibfnamefont{J.}~\bibnamefont{Damron}},
  \bibinfo{author}{\bibfnamefont{N.}~\bibnamefont{Ristoff}},
  \bibinfo{author}{\bibfnamefont{N.}~\bibnamefont{Mosavian}}, \bibnamefont{and}
  \bibinfo{author}{\bibfnamefont{V.~M.} \bibnamefont{Acosta}},
  \bibinfo{journal}{Physical Review Research} \textbf{\bibinfo{volume}{2}},
  \bibinfo{pages}{023394} (\bibinfo{year}{2020}), ISSN
  \bibinfo{issn}{2643-1564},
  \urlprefix\url{https://link.aps.org/doi/10.1103/PhysRevResearch.2.023394}.

\bibitem[{\citenamefont{Ziegler et~al.}(2010)\citenamefont{Ziegler, Ziegler,
  and Biersack}}]{ziegler_srim_2010}
\bibinfo{author}{\bibfnamefont{J.~F.} \bibnamefont{Ziegler}},
  \bibinfo{author}{\bibfnamefont{M.~D.} \bibnamefont{Ziegler}},
  \bibnamefont{and} \bibinfo{author}{\bibfnamefont{J.~P.}
  \bibnamefont{Biersack}}, \bibinfo{journal}{Nuclear Instruments and Methods in
  Physics Research Section B: Beam Interactions with Materials and Atoms}
  \textbf{\bibinfo{volume}{268}}, \bibinfo{pages}{1818} (\bibinfo{year}{2010}),
  ISSN \bibinfo{issn}{0168-583X},
  \urlprefix\url{http://www.sciencedirect.com/science/article/pii/S0168583X10001862}.

\bibitem[{\citenamefont{Berzins
  et~al.}(2021{\natexlab{a}})\citenamefont{Berzins, Smits, and
  Petruhins}}]{berzins_characterization_2021}
\bibinfo{author}{\bibfnamefont{A.}~\bibnamefont{Berzins}},
  \bibinfo{author}{\bibfnamefont{J.}~\bibnamefont{Smits}}, \bibnamefont{and}
  \bibinfo{author}{\bibfnamefont{A.}~\bibnamefont{Petruhins}},
  \bibinfo{journal}{Materials Chemistry and Physics}
  \textbf{\bibinfo{volume}{267}}, \bibinfo{pages}{124617}
  (\bibinfo{year}{2021}{\natexlab{a}}), ISSN \bibinfo{issn}{02540584},
  \urlprefix\url{https://linkinghub.elsevier.com/retrieve/pii/S0254058421004004}.

\bibitem[{\citenamefont{Alberts and Boeyens}(1976)}]{alberts_elastic_1976}
\bibinfo{author}{\bibfnamefont{H.}~\bibnamefont{Alberts}} \bibnamefont{and}
  \bibinfo{author}{\bibfnamefont{J.}~\bibnamefont{Boeyens}},
  \bibinfo{journal}{Journal of Magnetism and Magnetic Materials}
  \textbf{\bibinfo{volume}{2}}, \bibinfo{pages}{327} (\bibinfo{year}{1976}),
  ISSN \bibinfo{issn}{03048853},
  \urlprefix\url{https://linkinghub.elsevier.com/retrieve/pii/0304885376900445}.

\bibitem[{\citenamefont{Kirchner}(1964)}]{kirchner_thermal_1964}
\bibinfo{author}{\bibfnamefont{H.}~\bibnamefont{Kirchner}},
  \bibinfo{journal}{Progress in Solid State Chemistry}
  \textbf{\bibinfo{volume}{1}}, \bibinfo{pages}{1} (\bibinfo{year}{1964}), ISSN
  \bibinfo{issn}{00796786},
  \urlprefix\url{https://linkinghub.elsevier.com/retrieve/pii/0079678664900020}.

\bibitem[{\citenamefont{Kudielka}(1972)}]{kudielka_thermische_1972}
\bibinfo{author}{\bibfnamefont{H.}~\bibnamefont{Kudielka}},
  \bibinfo{journal}{Monatshefte fur Chemie} \textbf{\bibinfo{volume}{103}},
  \bibinfo{pages}{72} (\bibinfo{year}{1972}), ISSN \bibinfo{issn}{0026-9247,
  1434-4475}, \urlprefix\url{http://link.springer.com/10.1007/BF00912929}.

\bibitem[{\citenamefont{Hidnert}(1941)}]{hidnert_thermal_1941}
\bibinfo{author}{\bibfnamefont{P.}~\bibnamefont{Hidnert}},
  \bibinfo{journal}{Journal of Research of the National Bureau of Standards,
  Research Paper RP 1361} \textbf{\bibinfo{volume}{26}}, \bibinfo{pages}{11}
  (\bibinfo{year}{1941}),
  \urlprefix\url{https://nvlpubs.nist.gov/nistpubs/jres/26/jresv26n1p81_A1b.pdf}.

\bibitem[{\citenamefont{Jacobson and Stoupin}(2019)}]{jacobson_thermal_2019}
\bibinfo{author}{\bibfnamefont{P.}~\bibnamefont{Jacobson}} \bibnamefont{and}
  \bibinfo{author}{\bibfnamefont{S.}~\bibnamefont{Stoupin}},
  \bibinfo{journal}{Diamond and Related Materials}
  \textbf{\bibinfo{volume}{97}}, \bibinfo{pages}{107469}
  (\bibinfo{year}{2019}), ISSN \bibinfo{issn}{09259635},
  \urlprefix\url{https://linkinghub.elsevier.com/retrieve/pii/S0925963519303851}.

\bibitem[{\citenamefont{Berzins
  et~al.}(2021{\natexlab{b}})\citenamefont{Berzins, Smits, Petruhins, and
  Grube}}]{berzins_surface_2021}
\bibinfo{author}{\bibfnamefont{A.}~\bibnamefont{Berzins}},
  \bibinfo{author}{\bibfnamefont{J.}~\bibnamefont{Smits}},
  \bibinfo{author}{\bibfnamefont{A.}~\bibnamefont{Petruhins}},
  \bibnamefont{and} \bibinfo{author}{\bibfnamefont{H.}~\bibnamefont{Grube}},
  \bibinfo{journal}{Materials Chemistry and Physics}
  \textbf{\bibinfo{volume}{272}}, \bibinfo{pages}{124972}
  (\bibinfo{year}{2021}{\natexlab{b}}), ISSN \bibinfo{issn}{02540584},
  \urlprefix\url{https://linkinghub.elsevier.com/retrieve/pii/S0254058421007550}.

\bibitem[{\citenamefont{Broadway et~al.}(2020)\citenamefont{Broadway, Scholten,
  Tan, Dontschuk, Lillie, Johnson, Zheng, Wang, Oganov, Tian
  et~al.}}]{broadway_imaging_2020}
\bibinfo{author}{\bibfnamefont{D.~A.} \bibnamefont{Broadway}},
  \bibinfo{author}{\bibfnamefont{S.~C.} \bibnamefont{Scholten}},
  \bibinfo{author}{\bibfnamefont{C.}~\bibnamefont{Tan}},
  \bibinfo{author}{\bibfnamefont{N.}~\bibnamefont{Dontschuk}},
  \bibinfo{author}{\bibfnamefont{S.~E.} \bibnamefont{Lillie}},
  \bibinfo{author}{\bibfnamefont{B.~C.} \bibnamefont{Johnson}},
  \bibinfo{author}{\bibfnamefont{G.}~\bibnamefont{Zheng}},
  \bibinfo{author}{\bibfnamefont{Z.}~\bibnamefont{Wang}},
  \bibinfo{author}{\bibfnamefont{A.~R.} \bibnamefont{Oganov}},
  \bibinfo{author}{\bibfnamefont{S.}~\bibnamefont{Tian}}, \bibnamefont{et~al.},
  \bibinfo{journal}{Advanced Materials} \textbf{\bibinfo{volume}{32}},
  \bibinfo{pages}{2003314} (\bibinfo{year}{2020}), ISSN
  \bibinfo{issn}{0935-9648, 1521-4095},
  \urlprefix\url{https://onlinelibrary.wiley.com/doi/10.1002/adma.202003314}.

\bibitem[{\citenamefont{Froggatt and
  Moore}(1998)}]{froggatt_high-spatial-resolution_1998}
\bibinfo{author}{\bibfnamefont{M.}~\bibnamefont{Froggatt}} \bibnamefont{and}
  \bibinfo{author}{\bibfnamefont{J.}~\bibnamefont{Moore}},
  \bibinfo{journal}{Applied Optics} \textbf{\bibinfo{volume}{37}},
  \bibinfo{pages}{1735} (\bibinfo{year}{1998}), ISSN \bibinfo{issn}{0003-6935,
  1539-4522},
  \urlprefix\url{https://www.osapublishing.org/abstract.cfm?URI=ao-37-10-1735}.

\bibitem[{\citenamefont{Cai et~al.}(2005)\citenamefont{Cai, Xu, Kang, Gibart,
  and Beaumont}}]{cai_high-spatial-resolution_2005}
\bibinfo{author}{\bibfnamefont{D.}~\bibnamefont{Cai}},
  \bibinfo{author}{\bibfnamefont{F.}~\bibnamefont{Xu}},
  \bibinfo{author}{\bibfnamefont{J.}~\bibnamefont{Kang}},
  \bibinfo{author}{\bibfnamefont{P.}~\bibnamefont{Gibart}}, \bibnamefont{and}
  \bibinfo{author}{\bibfnamefont{B.}~\bibnamefont{Beaumont}},
  \bibinfo{journal}{Applied Physics Letters} \textbf{\bibinfo{volume}{86}},
  \bibinfo{pages}{211917} (\bibinfo{year}{2005}), ISSN
  \bibinfo{issn}{0003-6951, 1077-3118},
  \urlprefix\url{http://aip.scitation.org/doi/10.1063/1.1929866}.

\bibitem[{\citenamefont{Marshall
  et~al.}(2021{\natexlab{b}})\citenamefont{Marshall, Phillips, Turner, Ku,
  Zhou, Delegan, Heremans, Holt, and Walsworth}}]{marshall_scanning_2021}
\bibinfo{author}{\bibfnamefont{M.~C.} \bibnamefont{Marshall}},
  \bibinfo{author}{\bibfnamefont{D.~F.} \bibnamefont{Phillips}},
  \bibinfo{author}{\bibfnamefont{M.~J.} \bibnamefont{Turner}},
  \bibinfo{author}{\bibfnamefont{M.~J.~H.} \bibnamefont{Ku}},
  \bibinfo{author}{\bibfnamefont{T.}~\bibnamefont{Zhou}},
  \bibinfo{author}{\bibfnamefont{N.}~\bibnamefont{Delegan}},
  \bibinfo{author}{\bibfnamefont{F.~J.} \bibnamefont{Heremans}},
  \bibinfo{author}{\bibfnamefont{M.~V.} \bibnamefont{Holt}}, \bibnamefont{and}
  \bibinfo{author}{\bibfnamefont{R.~L.} \bibnamefont{Walsworth}},
  \bibinfo{journal}{arXiv:2103.08388 [cond-mat, physics:physics,
  physics:quant-ph]}  (\bibinfo{year}{2021}{\natexlab{b}}),
  \bibinfo{note}{arXiv: 2103.08388},
  \urlprefix\url{http://arxiv.org/abs/2103.08388}.

\bibitem[{\citenamefont{Moore}(2009)}]{moore_imaging_2009}
\bibinfo{author}{\bibfnamefont{M.}~\bibnamefont{Moore}},
  \bibinfo{journal}{Journal of Physics: Condensed Matter}
  \textbf{\bibinfo{volume}{21}}, \bibinfo{pages}{364217}
  (\bibinfo{year}{2009}), ISSN \bibinfo{issn}{0953-8984, 1361-648X},
  \urlprefix\url{https://iopscience.iop.org/article/10.1088/0953-8984/21/36/364217}.

\bibitem[{\citenamefont{Friel et~al.}(2009)\citenamefont{Friel, Clewes,
  Dhillon, Perkins, Twitchen, and Scarsbrook}}]{friel_control_2009}
\bibinfo{author}{\bibfnamefont{I.}~\bibnamefont{Friel}},
  \bibinfo{author}{\bibfnamefont{S.}~\bibnamefont{Clewes}},
  \bibinfo{author}{\bibfnamefont{H.}~\bibnamefont{Dhillon}},
  \bibinfo{author}{\bibfnamefont{N.}~\bibnamefont{Perkins}},
  \bibinfo{author}{\bibfnamefont{D.}~\bibnamefont{Twitchen}}, \bibnamefont{and}
  \bibinfo{author}{\bibfnamefont{G.}~\bibnamefont{Scarsbrook}},
  \bibinfo{journal}{Diamond and Related Materials}
  \textbf{\bibinfo{volume}{18}}, \bibinfo{pages}{808} (\bibinfo{year}{2009}),
  ISSN \bibinfo{issn}{09259635},
  \urlprefix\url{https://linkinghub.elsevier.com/retrieve/pii/S0925963509000156}.

\bibitem[{\citenamefont{Crisci et~al.}(2011)\citenamefont{Crisci, Baillet,
  Mermoux, Bogdan, Nesládek, and Haenen}}]{crisci_residual_2011}
\bibinfo{author}{\bibfnamefont{A.}~\bibnamefont{Crisci}},
  \bibinfo{author}{\bibfnamefont{F.}~\bibnamefont{Baillet}},
  \bibinfo{author}{\bibfnamefont{M.}~\bibnamefont{Mermoux}},
  \bibinfo{author}{\bibfnamefont{G.}~\bibnamefont{Bogdan}},
  \bibinfo{author}{\bibfnamefont{M.}~\bibnamefont{Nesládek}},
  \bibnamefont{and} \bibinfo{author}{\bibfnamefont{K.}~\bibnamefont{Haenen}},
  \bibinfo{journal}{physica status solidi (a)} \textbf{\bibinfo{volume}{208}},
  \bibinfo{pages}{2038} (\bibinfo{year}{2011}), ISSN \bibinfo{issn}{18626300},
  \urlprefix\url{http://doi.wiley.com/10.1002/pssa.201100039}.

\bibitem[{\citenamefont{{P. L. Hanley, I. Kiflawi and Andrew Richard
  Lang}}(1977)}]{p_l_hanley_i_kiflawi_and_andrew_richard_lang_topographically_1977}
\bibinfo{author}{\bibnamefont{{P. L. Hanley, I. Kiflawi and Andrew Richard
  Lang}}}, \bibinfo{journal}{Philosophical Transactions of the Royal Society of
  London. Series A, Mathematical and Physical Sciences}
  \textbf{\bibinfo{volume}{284}}, \bibinfo{pages}{329} (\bibinfo{year}{1977}),
  ISSN \bibinfo{issn}{0080-4614, 2054-0272},
  \urlprefix\url{https://royalsocietypublishing.org/doi/10.1098/rsta.1977.0012}.

\bibitem[{\citenamefont{Kempf et~al.}(2007)\citenamefont{Kempf, Vignal,
  Cailletaud, Oltra, Weeber, and Finot}}]{kempf_high_2007}
\bibinfo{author}{\bibfnamefont{D.}~\bibnamefont{Kempf}},
  \bibinfo{author}{\bibfnamefont{V.}~\bibnamefont{Vignal}},
  \bibinfo{author}{\bibfnamefont{G.}~\bibnamefont{Cailletaud}},
  \bibinfo{author}{\bibfnamefont{R.}~\bibnamefont{Oltra}},
  \bibinfo{author}{\bibfnamefont{J.~C.} \bibnamefont{Weeber}},
  \bibnamefont{and} \bibinfo{author}{\bibfnamefont{E.}~\bibnamefont{Finot}},
  \bibinfo{journal}{Philosophical Magazine} \textbf{\bibinfo{volume}{87}},
  \bibinfo{pages}{1379} (\bibinfo{year}{2007}), ISSN \bibinfo{issn}{1478-6435,
  1478-6443},
  \urlprefix\url{http://www.tandfonline.com/doi/abs/10.1080/14786430600928527}.

\bibitem[{\citenamefont{Chipaux et~al.}(2015)\citenamefont{Chipaux, Tallaire,
  Achard, Pezzagna, Meijer, Jacques, Roch, and
  Debuisschert}}]{chipaux_magnetic_2015}
\bibinfo{author}{\bibfnamefont{M.}~\bibnamefont{Chipaux}},
  \bibinfo{author}{\bibfnamefont{A.}~\bibnamefont{Tallaire}},
  \bibinfo{author}{\bibfnamefont{J.}~\bibnamefont{Achard}},
  \bibinfo{author}{\bibfnamefont{S.}~\bibnamefont{Pezzagna}},
  \bibinfo{author}{\bibfnamefont{J.}~\bibnamefont{Meijer}},
  \bibinfo{author}{\bibfnamefont{V.}~\bibnamefont{Jacques}},
  \bibinfo{author}{\bibfnamefont{J.-F.} \bibnamefont{Roch}}, \bibnamefont{and}
  \bibinfo{author}{\bibfnamefont{T.}~\bibnamefont{Debuisschert}},
  \bibinfo{journal}{The European Physical Journal D}
  \textbf{\bibinfo{volume}{69}}, \bibinfo{pages}{166} (\bibinfo{year}{2015}),
  ISSN \bibinfo{issn}{1434-6060, 1434-6079},
  \urlprefix\url{http://link.springer.com/10.1140/epjd/e2015-60080-1}.

\bibitem[{\citenamefont{Ukhina et~al.}(2020)\citenamefont{Ukhina, Dudina,
  Esikov, Samoshkin, Stankus, Skovorodin, Galashov, and
  Bokhonov}}]{ukhina_influence_2020}
\bibinfo{author}{\bibfnamefont{A.~V.} \bibnamefont{Ukhina}},
  \bibinfo{author}{\bibfnamefont{D.~V.} \bibnamefont{Dudina}},
  \bibinfo{author}{\bibfnamefont{M.~A.} \bibnamefont{Esikov}},
  \bibinfo{author}{\bibfnamefont{D.~A.} \bibnamefont{Samoshkin}},
  \bibinfo{author}{\bibfnamefont{S.~V.} \bibnamefont{Stankus}},
  \bibinfo{author}{\bibfnamefont{I.~N.} \bibnamefont{Skovorodin}},
  \bibinfo{author}{\bibfnamefont{E.~N.} \bibnamefont{Galashov}},
  \bibnamefont{and} \bibinfo{author}{\bibfnamefont{B.~B.}
  \bibnamefont{Bokhonov}}, \bibinfo{journal}{Surface and Coatings Technology}
  \textbf{\bibinfo{volume}{401}}, \bibinfo{pages}{126272}
  (\bibinfo{year}{2020}), ISSN \bibinfo{issn}{02578972},
  \urlprefix\url{https://linkinghub.elsevier.com/retrieve/pii/S0257897220309415}.

\bibitem[{\citenamefont{Duan et~al.}(2019)\citenamefont{Duan, Du, Kavatamane,
  Arumugam, Tzeng, Chang, and Balasubramanian}}]{duan_efficient_2019}
\bibinfo{author}{\bibfnamefont{D.}~\bibnamefont{Duan}},
  \bibinfo{author}{\bibfnamefont{G.~X.} \bibnamefont{Du}},
  \bibinfo{author}{\bibfnamefont{V.~K.} \bibnamefont{Kavatamane}},
  \bibinfo{author}{\bibfnamefont{S.}~\bibnamefont{Arumugam}},
  \bibinfo{author}{\bibfnamefont{Y.-K.} \bibnamefont{Tzeng}},
  \bibinfo{author}{\bibfnamefont{H.-C.} \bibnamefont{Chang}}, \bibnamefont{and}
  \bibinfo{author}{\bibfnamefont{G.}~\bibnamefont{Balasubramanian}},
  \bibinfo{journal}{Optics Express} \textbf{\bibinfo{volume}{27}},
  \bibinfo{pages}{6734} (\bibinfo{year}{2019}), ISSN \bibinfo{issn}{1094-4087},
  \urlprefix\url{https://www.osapublishing.org/abstract.cfm?URI=oe-27-5-6734}.

\end{thebibliography}

\end{document}